%% file: main_2.tex
\documentclass[manuscript,screen,nonacm]{acmart}
\setcopyright{none}
\usepackage{multirow}
\usepackage{makecell}
\usepackage{subcaption}
\usepackage{graphicx}
\usepackage{cleveref}
\usepackage{color}
\usepackage{float}
\usepackage{booktabs}
\AtBeginDocument{%
  }

\setcopyright{acmlicensed}
\copyrightyear{2018}
\acmYear{2018}
\acmDOI{XXXXXXX.XXXXXXX}
\acmConference[Conference acronym 'XX]{Make sure to enter the correct
  conference title from your rights confirmation email}{June 03--05,
  2018}{Woodstock, NY}
\acmISBN{978-1-4503-XXXX-X/2018/06}

\begin{document}

\title{Understanding Goalball Training Across Roles and Team Contexts: Practices, Perceptions, Challenges, and Technological Opportunities}

\author{Gyeongdeok Kim}
\affiliation{%
  \institution{Gwangju Institute of Science and Technology}
  \city{Gwangju}
  \country{Republic of Korea}
}
\email{kimkunduk@gm.gist.ac.kr}

\author{Junhyuk Jang}
\affiliation{%
  \institution{Gwangju Institute of Science and Technology}
  \city{Gwangju}
  \country{Republic of Korea}
}
\email{jangjh2004@gm.gist.ac.kr}

\author{Donghyeon Kim}
\affiliation{%
  \institution{Gwangju Institute of Science and Technology}
  \city{Gwangju}
  \country{Republic of Korea}
}

\author{Billy Park}
\affiliation{%
  \institution{Gwangju Institute of Science and Technology}
  \city{Gwangju}
  \country{Republic of Korea}
}

\author{Gunhyuk Park}
\affiliation{%
  \department{Haptic Assistive Media Laboratory}
  \institution{Gwangju Institute of Science and Technology}
  \city{Gwangju}
  \country{Republic of Korea}
}
\email{maharaga@gist.ac.kr}

\renewcommand{\shortauthors}{Kim et al.}

\renewcommand{\shortauthors}{Trovato et al.}

\begin{abstract}
Goalball is a team sport for visually impaired athletes, played by throwing and blocking a ball while wearing eyeshades.
However, existing studies provide limited guidance on which everyday training difficulties technology should address for players and coaches across team contexts. 
We conducted semi-structured interviews and focus groups with 18 players and 12 coaches from seven South Korean teams spanning youth, second-division, first-division, and national-team contexts. 
Using qualitative coding and matrix-based analysis, we describe current training practices, participants' views of these methods, the challenges they encounter, and their technological support needs.
Drawing on the relationships among these four areas, we discuss implications for the design and evaluation of future support technologies that fit everyday goalball training across roles and team contexts.
\end{abstract}





\maketitle
\input{1_Introduction}
\input{2_Related-work}
\input{3_Method}
\input{4_Findings}
\input{5_Discussion}

\input{6_Limitation}
\input{7_Conclusion}

\begin{acks}
Hidden for Anonymity.
\end{acks}

\bibliographystyle{ACM-Reference-Format}
\bibliography{99_BIBLIOGRAPHIES}

\appendix

\end{document}

%% file: 1_Introduction.tex
\section{Introduction}
Sport and physical activity play an important role in health, social connection, and enjoyment for disabled people~\cite{strobel2025hci}. 
However, people who are blind or have low vision (BLV) often face barriers to accessing physical activities because they usually require visual demonstration, observation, and imitation in learning, even in sports designed for BLVs~\cite{de2023understanding, tanaka2026touching}. 
Even with verbal explanation or physical guidance, instructors may struggle to convey spatial and movement details precisely, safely, and in ways learners can revisit~\cite{de2023understanding, tanaka2026touching}. 
These barriers are not only individual learning problems, but also reflect broader limitations in instruction, coaching resources, facilities, or participation pathways~\cite{kim2023forblind, choi2024female, de2023understanding}.

\begin{figure*}[t]
    \centering
    \includegraphics[width=\textwidth]{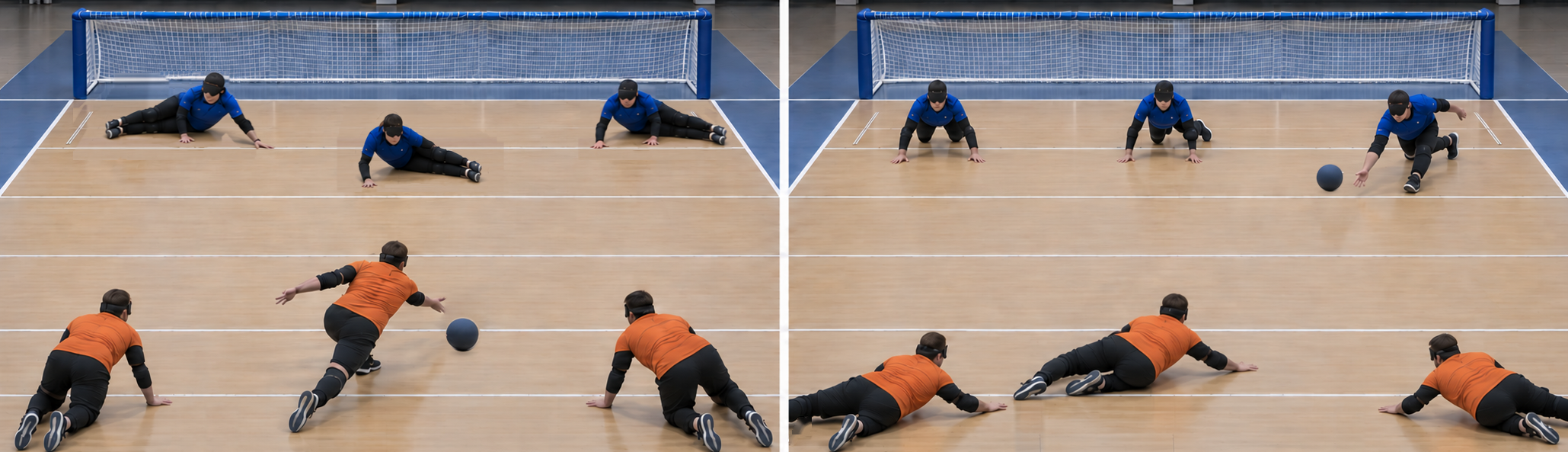}
    \caption{Illustrative image of the reciprocal attack--defense transition during goalball play. The orange team attacks while the blue team defends (left); after controlling the ball, the blue team attacks while the orange team reorganizes to defend (right).}
    % It was generated by GPT-5.6.}
    \Description{This image has two panels illustrating the transition between attack and defense in goalball. Each team has three players, all wearing eyeshades. In the left panel, an orange-team player throws the ball toward the blue team, whose three defenders are lying diagonally close to the ground to block it. In the right panel, the roles reverse: a blue-team player throws toward the orange team, whose three players have moved into defensive positions. In both panels, the center defender is positioned slightly closer to the opposing team than the two wing defenders.}
    \label{fig:goalball-play-transition}
\end{figure*}

%Goalball is one of the few sports specifically designed for BLV athletes with no able-bodied equivalent, and a rule system built around auditory and tactile play~\cite{bestebroer2025goalball}.
Goalball is designed specifically for BLV athletes rather than adapted from an able-bodied sport, making sound and tactile orientation central to play.
Two teams of three compete on an 18 $\times$ 9 m court using a ball with bells and tactile floor markings~\cite{bestebroer2025goalball, ibsa2025goalballrules, fujita2021ball}.
Teams typically field one center and two wings, with wings more attack-oriented and centers more defense-oriented in prior match analysis~\cite{fujita2021ball}.
During open play, teams alternate between attack and defense: a team has ten seconds from first defensive contact to return the ball across the center line and resumes defense immediately after throwing~\cite{bestebroer2025goalball,ibsa2025goalballrules,fujita2021ball} (Figure~\ref{fig:goalball-play-transition}).
Rules also restrict unnecessary noise during throws and coaching communication~\cite{ibsa2025goalballrules}.
Players wear eyeshades to equalize visual conditions during play and use sound, tactile markings, and spatial awareness to orient and respond.
In training, players must learn to orient themselves on the court, interpret the moving ball through sound, execute throwing and blocking actions.
% Simultaneously, they should respond quickly under strict rules governing noise and play, such as penalties for unnecessary noise during throws, restricted coaching communication, and a ten-second limit after defensive contact.
Coaches therefore need ways to teach orientation and technical skills that do not depend solely on visual demonstration or feedback~\cite{de2024coaching}.
Thus, goalball training is not simply a matter of physical repetition, but a process of aligning non-visual information, body movement, and team coordination through coaching and practice.

%Meanwhile, existing goalball research has provided important knowledge about performance, participation, and coaching. 
Existing research has examined several aspects of goalball and related non-visual sport practices.
Goalball research has examined performance factors~\cite{woo2018goalball, molik2015game}, imagery, participation, and coaching experiences~\cite{eddy2003mental, antritsou2016socialisation, choi2016sports, de2024coaching, macdonald2020you}.
Para-sport coaching studies further emphasize adaptive, athlete-centered instruction and supportive coach–athlete relationships~\cite{mycock2020blind, ovstehage2025exploring, ovstehage2026coaches}.
Other work explores technologies for nonvisual movement representation, accessible spatial and remote learning, and adaptive-sport analytics~\cite{khurana2021beyond, de2023understanding, tanaka2026touching}, alongside enjoyment, community, and disabled athletes' values~\cite{strobel2025hci}.

However, these studies remain limited in two ways.
First, they tend to examine performance, experience, coaching, or technology as separate topics, rather than explaining how these dimensions come together in everyday goalball training.
% Second, their focus on particular populations or perspectives also leaves limited understanding of differences in training practices and needs between players and coaches and across team contexts.
Second, their focus on particular populations or perspectives leaves limited understanding of how players and coaches experience everyday training and identify support needs across team contexts.

We therefore ask:
\begin{itemize}
    \item RQ1. How are current goalball training practices organized and conducted across roles and team contexts?
    \item RQ2. How do players and coaches perceive current training methods, including the information provided and the ways communication occurs during training?
    \item RQ3. What challenges and breakdowns do players and coaches experience in current goalball training?
    \item RQ4. What technical opportunities emerge from these practices, perceptions, and challenges, and under what conditions could technology meaningfully support goalball training?
\end{itemize}

To answer these \textit{research questions}, we conducted semi-structured individual interviews and focus groups with goalball players and coaching staff in South Korea.
The Korean goalball ecosystem is well suited to this study because it brings together multiple interconnected training contexts within a single national system, including youth teams, second-division teams, first-division business teams, and national teams. 
We analyzed the data through an iterative qualitative coding process, moving from an initial codebook to refined themes that capture training practices, perceptions, challenges, and technical opportunities across roles and team contexts.
After finalizing the codebook, we used participants as cases in an additional matrix-based analysis to examine patterns and relationships across the coded data~\cite{gale2013using, miles2013qualitative}.

Our findings show 1) how individual training addresses physical and sensory demands, 2) how team-based practice combines tactical coordination and communication, 3) how performance review is conducted and where access remains difficult, and 4) how training is organized around goals and available resources. 
Across these four domains, we connect current practices with participants' perceptions and challenges to identify technical requirements. 
We then discuss 1) accessible learning across vision conditions and histories, 2) essential feedback that explains why outcomes occur and what to change next, 3) the spatial coordination burden concentrated on centers and ways to share it among teammates, and 4) technological support aligned with teams' goals and resources.
Together, these analyses provide an empirical basis for designing and evaluating technologies grounded in everyday goalball training from multifaceted populations.
% Through this study, we contribute a qualitative analysis of how goalball training is organized, communicated, and experienced across roles and competitive levels, and we identify design considerations for accessible training technologies grounded in these everyday practices.

%% file: 2_Related-work.tex
\section{Related Work}
Prior research on goalball, para-sport training, and accessible movement technologies contextualizes our study of everyday training practices and technology needs across roles and team contexts.

\subsection{Goalball Play and Sporting Experience}
\subsubsection{Performance and Tactical Factors}
Goalball game-analysis research associates offensive effectiveness with player position, throw execution, target area, trajectory, and ball travel time~\cite{woo2018goalball, link2018finding, morato2018faster}.
Among elite male players, some offensive and defensive indicators differed by degree of visual impairment, while body-size measures did not clearly distinguish performance~\cite{molik2015game}.
A systematic review of goalball sports preparation further identifies physical conditioning, throwing ability, role-specific demands, and psychological preparation as factors relevant to performance development~\cite{fernandez2021sports}. 
Together, these studies clarify how actions and player characteristics may relate to goalball performance.

\subsubsection{Participation, Experience, and Coaching}
Beyond performance, research has examined how athletes enter, sustain, and make sense of goalball participation.
Antritsou describes participation as shaped by schooling and physical education, supported by friends, family, and organizations, and sustained through relationships within the goalball community~\cite{antritsou2016socialisation}.
Choi et al. report tensions between school and sport, uncertainty about future athletic careers, and new relationships and life opportunities among student athletes~\cite{choi2016sports}, while MacDonald et al. describe how collegiate goalball can support team cohesion, enjoyment, pride, and changed understandings of disability~\cite{macdonald2020you}.
Eddy and Mellalieu show how elite goalball athletes rehearse skills, strategies, and competitive situations through auditory, tactile, spatial, and kinesthetic imagery~\cite{eddy2003mental}, highlighting the internal representation of movement, space, sound, and competition.

Coaching-focused research adds another layer to this picture. 
Using the Coaching Behavior Scale for Sport~\cite{cote1999exploratory}, De Luna-Vazquez et al. found that goalball athletes generally rated their coaches’ behaviors positively, with higher scores for personal rapport, technical skill instruction, and competition strategies and lower scores for physical training and goal setting~\cite{de2024coaching}.
Research on youth goalball coaches identifies a lack of sport-specific coaching manuals and guides and a need for formal coach development programs~\cite{esatbeyouglu2024coaching}.
Historical work based on a goalball coach's oral history further emphasizes the need for long-term approaches to player development and athlete support~\cite{lee2022goalball}.
Research on blind football and para-sport coaching similarly emphasizes adaptive, athlete-centered approaches, supportive coach–athlete relationships, and regular coach–athlete communication in high-quality training~\cite{mycock2020blind, ovstehage2025exploring, ovstehage2026coaches}.

Together, these studies show that goalball participation and training are shaped by social pathways, sensory experience, coaching relationships, and sport-specific support. 
Our study examines how these dimensions intersect in everyday training and inform support needs across player and coaching roles and team contexts.

%\subsection{Training Quality and Coaching in Visual Impairment and Other Para-Sport}

\subsection{Accessible Movement Learning and Sports Technology}
Human–computer interaction (HCI) and accessibility research has increasingly examined how technology can support movement learning and adaptive sport beyond goalball.
De Silva et al. identify design challenges in BLV movement education involving movement representation, feedback, kinesthetic awareness, spatial and social learning, and accessible remote learning~\cite{de2023understanding}.
Khurana et al. identify opportunities for improving accessibility and data-driven support through adaptive-sport analytics~\cite{khurana2021beyond}.
Other HCI work has demonstrated different approaches for supporting accessible movement and sport experiences, including adaptive kickboxing equipment for people with low vision~\cite{tsutsui2024low} and 3D tactile poses for blind people's movement learning~\cite{tanaka2026touching}.
Strobel and Gerling caution against exclusively health-oriented or replacement-based framings, advocating attention to leisure, enjoyment, community, interdependence, and disabled people’s values~\cite{strobel2025hci}.

Compared with broader sports technology research, goalball-specific work remains relatively limited.
GoalBaural uses acoustic virtual reality for off-court practice in locating thrown balls, with evaluation results suggesting improvements in the speed and accuracy of locating balls~\cite{miura2018goalbaural}.
Performance-analysis tools have also been proposed to support match data collection, visualization, and measurement of characteristics such as court sectors, defensive position, ball direction, and speed~\cite{weber2015performance}.
These systems suggest benefits from sport-specific technical support but primarily target bounded tasks, such as acoustic perception training or performance analysis.

Less is known about how goalball technology needs emerge from training routines, coach–player communication, tactical learning, feedback, stakeholder perceptions, and practical challenges.
Our study examines these connections across roles and team contexts.

%% file: 3_Method.tex
\section{Methods}
Our qualitative study examined goalball training practices, experiences, challenges, and technology needs through focus groups and semi-structured interviews.

\subsection{Focus Groups and Semi-structured Interviews}
% To understand everyday goalball training practices and technology needs, we engaged goalball players and coaching staff through focus groups and semi-structured interviews.
% We selected South Korea as our empirical site because its goalball ecosystem includes multiple interconnected team contexts within a single national system, including youth, second-division, first-division, and national teams. 
% This setting allowed us to examine goalball training as situated within a broader sport ecosystem rather than as isolated practices of a single team or participant group.
We selected South Korea because its goalball ecosystem includes interconnected youth, second-division, first-division, and national-team contexts within one national system.
Its national programs were also internationally competitive: South Korea ranked 10th among 62 women’s teams and 15th among 100 men’s teams in the June 2026 International Blind Sports Federation (IBSA) world rankings~\cite{ibsaGoalballWomensRankings2026,ibsaGoalballMensRankings2026}.
This combination of multiple domestic team levels and internationally competitive national-team contexts allowed us to examine goalball training as situated within a broader sport ecosystem rather than as isolated practices of a single team or participant group.

Where group discussion was feasible, we held separate focus groups for players and coaching staff within each team.
This format allowed participants sharing roles and team contexts to reflect on training routines, compare experiences, and elaborate on each other's comments.
When only one participant from a role group was available, or team and individual schedules prevented group discussion, we conducted individual semi-structured interviews.
Both formats used the same core topics, open-ended questions, and follow-up discussion.

The discussions focused on four areas aligned with our \textit{research questions}: 1) current training practices, 2) perceptions of training and feedback, 3) training-related challenges, and 4) technical needs and design considerations for future support. 
We encouraged participants to describe concrete examples from training and competition preparation, such as feedback exchange, tactical understanding, physical and sensory training, training and resource management, perceived benefits and limitations, practical challenges, and possible technological supports.

\begin{table*}[t]
\centering

\caption{Participant demographics, team context, and interview method.}
\label{tab:participants}

{\renewcommand{\arraystretch}{0.92}
\resizebox{\textwidth}{!}{%
\begin{tabular}{cccccc}
\toprule
\textbf{ID}\textsuperscript{a} &
\textbf{Gender} &
\textbf{Team Context}\textsuperscript{b} &
\textbf{Vision Condition} &
\textbf{Current Role} &
\textbf{Interview Method}\textsuperscript{c} \\
\midrule

Y1P1 & Man & Y1 (Youth) & Blind (inborn) & Player (Wing) & TI \\
Y1C1 & Man & Y1 (Youth) & No visual impairment & Coach (Head) & II \\

Y2C1 & Man & Y2 (Youth) & No visual impairment & Coach (Head) & II \\
Y2C2 & Man & Y2 (Youth) & Low vision (inborn) & Coach (Assistant) & FG \\
Y2C3 & Man & Y2 (Youth) & No visual impairment & Coach (Assistant) & FG \\

\midrule

S1P1 & Woman & S1 (Second division) & Low vision (acquired) & Player (Wing) & II \\
S1P2 & Man & S1 (Second division) & Blind (inborn) & Player (Both) & II \\
S1P3 & Man & S1 (Second division) & Blind (acquired) & Player (Wing) & II \\
S1P4 & Man & S1 (Second division) & Low vision (acquired) & Player (Wing) & II \\
S1C1 & Woman & S1 (Second division) & No visual impairment & Coach (Head) & II \\
S1C2 & Man & S1 (Second division) & No visual impairment & Coach (Assistant) & II \\

S2P1 & Man & S2 (Second division) & Low vision (inborn) & Player (Both) & FG \\
S2P2 & Man & S2 (Second division) & Blind (inborn) & Player (Center) & FG \\
S2P3 & Man & S2 (Second division) & Low vision (acquired) & Player (Wing) & FG \\

\midrule

F1P1 & Man & F1 (First division) & Low vision (acquired) & Player (Both) & FG \\
F1P2 & Man & F1 (First division) & Low vision (inborn) & Player (Wing) & FG \\
F1P3 & Man & F1 (First division) & Low vision (acquired) & Player (Center) & FG \\
F1C1 & Man & F1 (First division) & No visual impairment & Coach (Head) & II \\

F2P1 & Man & F2 (First division) & Low vision (inborn) & Player (Center) & FG \\
F2P2 & Man & F2 (First division) & Blind (inborn) & Player (Wing) & FG \\
F2C1 & Man & F2 (First division) & No visual impairment & Coach (Head) & FG \\
F2C2 & Man & F2 (First division) & No visual impairment & Coach (Assistant) & FG \\

\midrule

N1P1 & Woman & N1 (National) & Low vision (inborn) & Player (Center) & FG \\
N1P2 & Woman & N1 (National) & Low vision (inborn) & Player (Wing) & FG \\
N1P3 & Woman & N1 (National) & Low vision (inborn) & Player (Wing) & FG \\
N1P4 & Woman & N1 (National) & Blind (inborn) & Player (Center) & FG \\
N1P5 & Woman & N1 (National) & Low vision (acquired) & Player (Both) & FG \\
N1C1 & Woman & N1 (National) & No visual impairment & Coach (Head) & FG \\
N1C2 & Man & N1 (National) & Low vision (inborn) & Coach (Assistant) & FG \\
N1C3 & Woman & N1 (National) & No visual impairment & Coach (Assistant) & FG \\

\bottomrule
\end{tabular}%
}
}

\vspace{3pt}

\begin{minipage}{\textwidth}
\scriptsize
\raggedright

\textsuperscript{a}
IDs combine team context, current role, and participant number.
P indicates player and C indicates coach.

\textsuperscript{b}
Y indicates youth, S second division, F first division, and N national team.
Numerals distinguish teams within the same competitive level.

\textsuperscript{c}
FG indicates focus group, II individual semi-structured interview, and TI individual telephone interview.

\end{minipage}

\Description{
Participant characteristics for 18 players and 12 coaches from seven South Korean goalball teams.
Rows are grouped by youth, second-division, first-division, and national-team contexts, with players listed before coaches within each team.
From left to right, the columns report participant ID, gender, team context, vision condition and history, current role, and interview format.
}

\end{table*}

\subsection{Participants}
We recruited 30 participants, comprising 18 players and 12 coaching staff, from seven South Korean teams: two youth, two second-division, two first-division, and one women’s national team.
With support from the Korea Blind Goalball Federation (KBGF), we coordinated recruitment, scheduling, and locations with each team’s coaching staff.
Five teams included both roles. 
One youth team included only coaching staff because players declined due to training-related fatigue and perceived burden, while one second-division team included only players.
The authors were external researchers and held no player or coaching roles.% within the participating goalball teams.

Sessions accommodated each team’s availability and used familiar or accessible settings, including training facilities and an accommodation lounge. 
Most sessions were conducted in person, although one youth player participated in a telephone interview because an in-person group session was infeasible.
Participants were compensated at approximately \$14.5 USD per hour.

Table~\ref{tab:participants} reports gender, team context, vision condition, current role, and interview format.
Players' current roles included wing, center, or both positions, and coaching roles included head and assistant coach.
Participants ranged from 18 to 55 years of age ($M=31.2$), with 21 men and 9 women.
The sample included 5 participants from youth teams, 9 from second-division teams, 8 from first-division teams, and 8 from the women's national team.
Vision conditions comprised 6 blind participants, 14 participants with low vision, and 10 participants without visual impairment.
Current players reported 1 to 25 years of playing experience, and current coaching staff reported 2 to 15 years of coaching experience.
Twenty participants took part in focus groups and ten took part in individual interviews, including one telephone interview.
% When an interviewee had experience on both coaching and playing, we retained both values in the table.
Although we collected additional contextual information to support analysis, we report only what is needed to interpret the findings, omitting details including ages and years of playing or coaching experience to prevent potential identification of individuals.

\subsection{Data Collection and Analysis}
The study was approved by the university's Institutional Review Board (anonymized for review).
Before each session, participants provided informed consent to participate in the study and verbally agreed to recording.
We conducted 7 focus groups and 10 individual semi-structured interviews, yielding 17 transcripts.
Individual interviews averaged 58 minutes, and focus groups averaged 107 minutes, with the longest lasting 150 minutes.
Plaud AI~\cite{plaudai} automatically transcribed the recordings.
The authors who conducted each session checked the transcripts against the recordings and corrected transcription errors.
All sessions were conducted and analyzed in Korean, and quotations were translated into English during manuscript preparation.

We analyzed the transcripts through open coding~\cite{charmaz2006constructing} and iterative codebook development, with four authors involved in the interviews serving as coders.
To develop a preliminary codebook, the first author served as the lead coder and initially open-coded four transcripts. 
The four \textit{research questions} provided the top-level analytic frame, while the lower levels of the codebook were developed inductively from the data. 
At these lower levels, the lead coder generated excerpt-level \textit{open codes} from transcript segments, consolidated similar \textit{open codes} into \textit{merged codes}, and organized related \textit{merged codes} into broader \textit{categories} under each \textit{research question}.
The preliminary codebook was then refined through peer review with the other coders. 
During this review process, the coders discussed the code hierarchy, code definitions, and excerpt-to-code assignments, and revised the codebook by splitting overly broad codes, merging overlapping codes, renaming unclear codes, reassigning excerpts to better-fitting codes, deleting redundant or unsupported codes, and adding new codes when the data introduced concepts not yet captured by the codebook.

After developing the preliminary codebook, the four coders independently coded one shared transcript and compared decisions to align code interpretations, naming conventions, excerpt boundaries, and assignments.
It helped establish shared guidelines for subsequent coding, peer review, and revision as the codebook evolved across the remaining twelve transcripts.
The remaining transcripts were then allocated among the four coders using two strategies: overlapping coding for paired review ($n=5/12$) and single-coder coding ($n=7/12$).
For overlapping coding, selected transcripts were independently coded by two coders, who then compared their coding and reconciled differences through discussion.
For single-coder coding, the other transcripts were assigned to one primary coder.

All subsequent coding remained open to the same peer-review practices used for the preliminary codebook.
However, in these later reviews, the coders also examined related lower-level codes and associated excerpts to determine whether the affected part of the codebook needed structural revision, including adjustments to code boundaries, grouping, or hierarchical relationships.
Throughout the analysis, the coders developed a shared terminology record for agreed definitions of codes and key analytic terms, along with coding rules and distinctions between similar concepts. 
This record was updated whenever ambiguities or needed distinctions emerged.
We used NVivo~\cite{nvivo} to manage transcripts, coded excerpts, codebook revisions, terminology notes, and the evolving code hierarchy.
% Through this process, the final codebook retained the four \textit{research questions} as its top-level analytic frame, linking the inductively developed and revised \textit{categories}, \textit{merged codes}, and \textit{open codes}. 
Through this process, the final codebook comprised 4 \textit{research questions}, 20 \textit{categories}, 80 \textit{merged codes}, and 953 \textit{open codes}, organized in a four-level hierarchy.
The resulting hierarchy linked granular transcript excerpts to recurring concepts and broader categories, allowing us to preserve participants' fine-grained accounts.

\begin{figure*}[t]
    \centering
    \includegraphics[width=\textwidth]{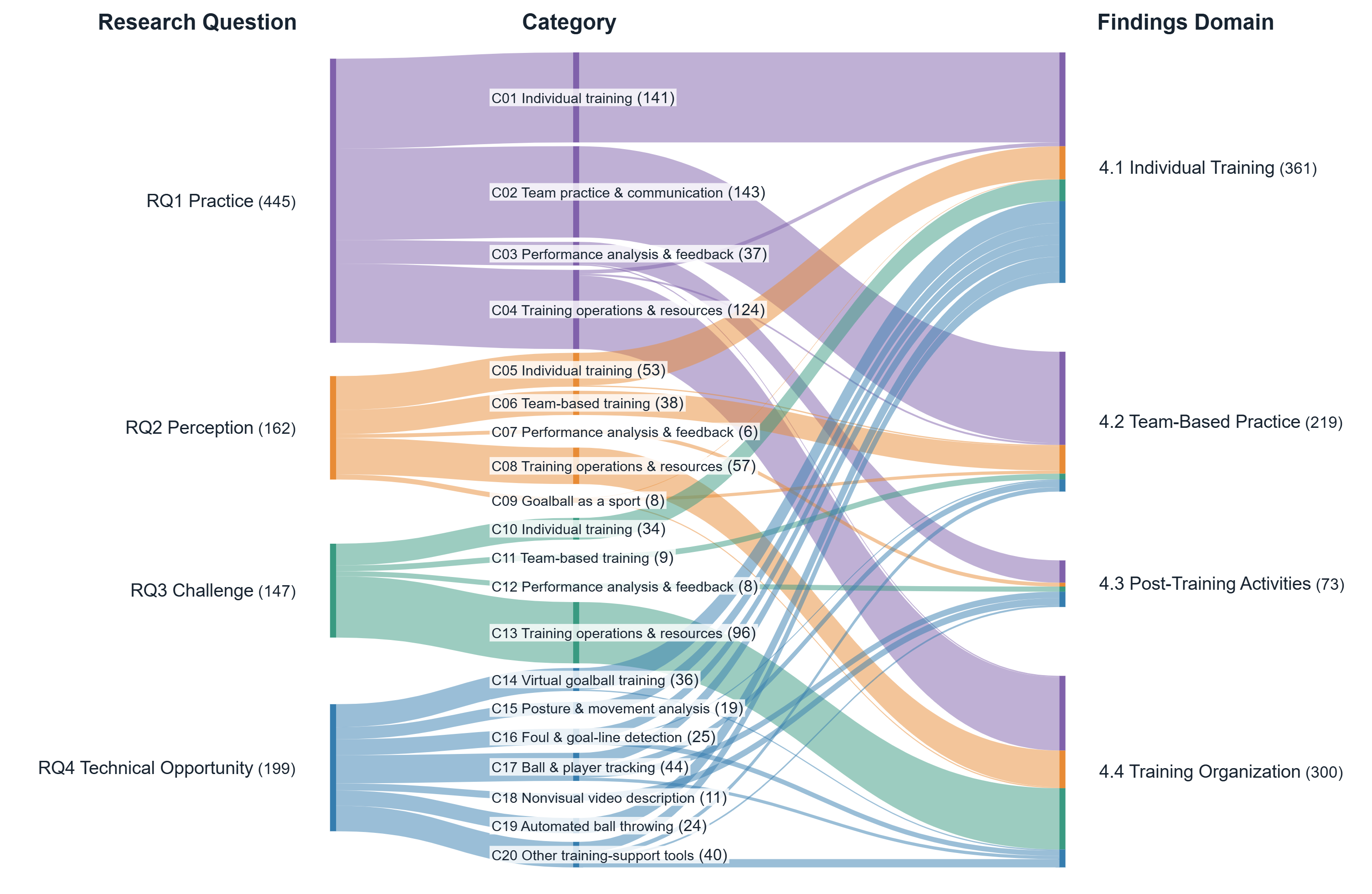}
    \caption{Connections between the codebook's two upper levels, \textit{research questions} and \textit{categories}, and the four findings domains used in the subsequent analysis, whose results are reported in (Section~\ref{sec:findings}). Ribbon widths and numbers indicate \textit{open code} counts.}
    \label{fig:codebook-domains}
    \Description{The left column organizes 953 open codes into four research-question groups. There are 445 codes describing practices, 162 codes concerning perceptions, 147 codes identifying challenges, and 199 codes concerning technical opportunities. The middle column groups these codes into 20 categories under their respective research questions. Connections to the right column show how these categories contribute to four findings domains. Individual training brings together 361 open codes, while team-based practice brings together 219 open codes. Post-training activities contains 73 open codes, and training organization contains 300 open codes. Each findings domain combines codes from all four research-question groups. Wider connecting bands indicate that more open codes follow that connection.}
\end{figure*}

We then conducted an additional matrix-based analysis to examine patterns and relationships across the coded data, treating each participant as a case and demographic characteristics as case-level attributes.
Drawing on the Framework Method's case-by-code organization~\cite{gale2013using} and Miles et al.'s guidance on qualitative data displays~\cite{miles2013qualitative}, we used linked matrix-based displays to compare the coded data across cases and examine patterns, contrasts, and relationships.
Using NVivo's Framework Matrix functionality, we organized the lowest-level \textit{open codes} by participant while retaining the broader \textit{merged codes}, \textit{categories}, and \textit{research questions} as hierarchical context and the coded excerpts as the evidentiary basis.
% We then linked the excerpts with participant metadata, including team context, current role, and vision condition, inventoried the \textit{open codes} across four broad domains of individual training, team-based practice, post-training performance review, and training operations and support, and regrouped the material around focal training phenomena.
% Figure~\ref{fig:codebook-domains} shows the connections from the codebook's two upper levels, \textit{research questions} and \textit{categories}, to these four findings domains.
% Candidate patterns, contrasts, and relationships were first recorded and then checked against original excerpts and contexts rather than inferred from labels or metadata alone.
% Candidates were retained as confirmed relationships only when excerpt-level comparison supported a coherent empirical connection.
We then linked the excerpts with participant metadata, including team context, current role, and vision condition, and organized the \textit{open codes} into four broad domains: individual training, team-based practice, post-training activities, and training organization (Figure~\ref{fig:codebook-domains}).
This organization helped us locate related accounts across the research questions, which we then regrouped around focal training phenomena for comparison.
Candidate patterns, contrasts, and relationships identified through these comparisons were recorded and then verified against the original excerpts and their contexts rather than inferred from code labels or metadata alone.
This verification process produced a relational display containing 112 confirmed relationships and a second display organizing them into 14 higher-level empirical syntheses within and across the four domains, which informed the structure of the findings presented in Section~\ref{sec:findings-individual-training}.
The finalized codebook and both analytic displays are provided in the supplementary materials.

%% file: 4_Findings.tex
\section{Findings}
\label{sec:findings}
The findings are organized into four domains: individual training, team-based practice, post-training performance review, and training operations and support.
Within each domain, we present current practices, participants' perceptions and challenges, and the technical requirements identified by participants.

\subsection{Individual Training: Physical Conditioning, Technique Refinement, and Sensory Learning}
\label{sec:findings-individual-training}
Individual training encompassed three connected areas: physical conditioning adapted to players' bodies and positions, technique refinement through feedback on outcomes and movement execution, and sensory learning through auditory, tactile, and proprioceptive experience.

\subsubsection{Adapting physical conditioning to individual and positional demands}
\label{sec:findings-conditioning}
Across all seven teams, 15 players and 12 coaches collectively described three aspects of physical conditioning: adaptations to players' bodies and positions, its contribution to sustained goalball performance, and technical requirements for accessible guidance and measurement.

% Evidence path: B01, SYN001, REL001, REL015, REL017, REL018, REL087, REL088, REL089, REL063
% Supporting references: REF0114 (Y1C1), REF0384 (S1P3), REF0603 (F2P2), REF0079 (N1P4), REF0553 (S1P1).
\textbf{Conditioning adaptations.}
Twenty-five participants across all seven teams described physical conditioning practices, including adaptations for vision-related safety and the physical demands of different playing positions.
In this paper, \textit{physical conditioning} refers to warm-ups and strength, cardiovascular, and position-specific exercise intended to support physical readiness for goalball.
Y1C1 explained why his team primarily used fixed weight machines rather than free weights: \textit{``Because the players have visual impairments, the ordinary things we call bodyweight exercise and lifting barbells or dumbbells can be a bit dangerous, so we mostly use machines.''}
S1P3 described center training as emphasizing rapid defensive reactions and keeping blocked balls within reach.
For wings, S1P3 and F2P2 described repeated throwing and rapid returns to a side defensive position, which F2P2 described as physically demanding.
Reflecting on her move from center to an attacking role, S1P1 similarly reported that attacking required more stamina and energy than playing center.
%These equipment choices and position-specific drills varied across teams and did not form a single standardized routine for \textit{physical conditioning}.
% Overall, participants described adapting physical conditioning to players' access needs and positional responsibilities.

% Evidence path: B02, SYN001, REL002, REL003, REL004, REL008, REL062, REL063.
% Supporting references: REF0460 (N1C2), REF0452 (S1P4), REF0450 (Y2C3), REF0983 (F2P2), REF0449 (Y2C1).
\textbf{Sustaining match performance.}
Twelve participants from six teams (Y1, Y2, S1, F1, F2, and N1) linked physical conditioning to sustaining performance as fatigue accumulated during a match and highlighted the need for exercises tailored to goalball movements.
N1C2 described a fatigued player returning too slowly after a moving attack and leaving the team exposed to a fast counterattack, while Y2C1 explained that louder breathing during fatigue made the bells inside an approaching ball harder to hear.
S1P4 emphasized additional strength exercise as a basis for sustaining throwing performance, while Y2C3 emphasized conditioning exercises built around goalball movements.
F2P2 further explained why conditioning needed to reflect the demands of a particular sport: \textit{``A marathon runner may have better endurance than almost anyone, but that is different from running on a soccer field. Ultimately, to play goalball well, you need fitness specific to goalball.''}

% Evidence path: B03, SYN001, REL006, REL007, and REL106.
% Supporting references: REF0491 and REF0945 (Y1C1), REF0943 (S1C2), REF0927, REF0929, and REF0930 (Y2C1), and REF0944 (F1P1, F1P2, and F1P3).
% Direct quotation: REF0945 (Y1C1).
\textbf{Technical requirements for accessible conditioning.}
Six participants from four teams (Y1, Y2, S1, and F1) described two technical requirements for physical conditioning: accessible guidance during exercise and measurements for monitoring physical changes.
When players exercised without a coach nearby, Y1C1 requested equipment-based guidance: \textit{``If there were auditory cues, or tactile cues where they grip the equipment, they could exercise more accurately and with proper form.''}
S1C2 requested audio-based position guidance for players running independently, while Y1C1 and Y2C1 described a need to track changes in strength, balance, and physical condition over time.
% Three F1 players expanded a researcher-proposed muscle-monitoring concept by relating objective measurements to self-checking and potential injury prevention.
Three F1 players additionally suggested to use the muscle-monitoring concept, which was introduced by the authors, to self-checking and potential injury prevention.
% Together, these requests focused on accessible guidance during exercise and measurements that players could use to monitor physical changes.
% The scheduling and facility constraints on shared conditioning are examined in Section~\ref{sec:findings-operations}.

\subsubsection{Making performance outcomes and movement processes inspectable}
\label{sec:findings-process-feedback}
Across all seven teams, 16 players and 10 coaches collectively described how throwing and defensive outcomes were communicated during individual practice, why information about ball and body movement was also needed, and which technologies they requested to make both forms of information accessible.

% Evidence path: B04, SYN002, REL009, SYN004, REL020.
% Supporting references: REF0002 (S1C2), REF0005 (F1P3), REF0012 (S1C2), REF0019 (F1C1), REF0036 (N1C1), REF0040 (S2P3), REF0044 (S1C2), REF0058 (S1P3), REF0085 (S1P4), REF0103 (S1P1), REF0483 and REF0485 (S1P1), REF0495 (S1P3), and REF0496 (S1P4).
% Direct quotation: REF0012 (S1C2).
\textbf{Performance-outcome information.}
% Eight participants from four teams (S1, S2, F1, and N1) described how they receive feedback on the observable results of individual throws and defense.
% In this paper, \textit{performance-outcome information} refers to the observable results conveyed through this feedback.
Eight participants from four teams (S1, S2, F1, and N1) described receiving feedback on the observable results of individual throws and defensive actions, which we term \textit{performance-outcome information}.
Collectively, the participants described a throw's direction and numbered arrival zone, and whether it resulted in a foul.
For example, S1P3 described coaches providing feedback on whether a throw resulted in a high-ball or long-ball foul.
For defense, S1P1 and S1P4 described feedback identifying where a conceded ball had passed between defenders.
The number and width of the goal-based target zones varied across teams.
S1C2 explained that the coaching staff provided feedback on the actual zone and its direction relative to the target but left the next adjustment to the player: \textit{``That is up to the player. We give the information, and then the player has to adjust it.''}
S1P1, who is a winger, explained that, after being told that a conceded ball had passed between her and the center, she could focus on that area and extend her defensive coverage.
% Across these accounts, \textit{performance-outcome information} provided a concrete reference for adjusting a subsequent throw or defensive action.

% Evidence path: B05, SYN002, REL011, REL021, REL022, REL028.
% Supporting references: REF0613 (N1P1), REF0614 (F2P1), REF0620 and REF0629 (F2P2), REF0633 (Y2C3), and REF1008 (F1P1).
% Direct quotation: REF0613 (N1P1).
\textbf{Need for performance-process information.}
Eighteen participants from six teams (Y1, Y2, S1, F1, F2, and N1) described situations in which \textit{performance-outcome information} was available but did not explain why the result of a throw or defensive action had occurred.
In this paper, \textit{performance-process information} refers to information about how the ball and the player's body moved, such as the ball's travel path and the player's stepping, posture, timing, release, and contact with the ball during defense.
Eleven of these participants were players who described being unable to inspect their own movements under eyeshades.
N1P1 explained, \textit{``Because goalball is played with the eyes completely covered, we cannot accurately recognize our own movement paths or actions. That is the most difficult part, and it is why we come to rely on feedback.''}
F1P1 similarly distinguished knowing a ball's starting and arrival points from knowing the path that it had traveled.
F2P1 described the difficulty of making immediate judgments when facing fast throws and emphasized that a conceded goal could involve several factors rather than a single error.
The need for process information also arose when the movement a coach intended differed from what a player actually performed.
% F2P2 described coaches instructing players to throw between the opposing left wing and center, but explained that repeatedly telling the players that the ball had still arrived in the middle did not reveal why it had deviated from the intended course.
% F2P2 described the inclarity of coach's feedback, for example, throwing to the location between the opposing left wing and center was feedbacked as the ball falls in the middle and he did not know how to adjust his motions to throw accurately.
Y2C3 described difficulty providing feedback on unsuccessful execution because it was unclear whether the problem stemmed from how the instruction was communicated or from the player's difficulty carrying out the instructed movement. 
In a separate interview, F2P2 described the similar difficulty in interpreting coaches' instruction.
For example, multiple practices targeting the location between the opposing left wing and center led to falls in the middle repeatedly, where the coaches' feedback did not help adjust his motions to throw accurately.
%Together, these accounts showed that participants needed \textit{performance-process information} to examine whether an unexpected outcome was related to the ball's path, the player's movement, or the way an intended action had been communicated.

% Evidence path: B06, SYN002, REL011, REL028, REL029, REL033, REL066, REL103.
% Supporting references: REF0808 (Y1C1), REF0815 (S1P4), and REF0844 (N1P3).
\textbf{Technical requirements for performance feedback.}
Twenty-two participants across all seven teams requested technologies that could provide immediate \textit{performance-outcome information} as well as more detailed \textit{performance-process information} through nonvisual feedback during training.
Nine participants across all seven teams requested automatic audio feedback on target-zone and foul results immediately after each throw.
Ten participants from five teams (Y1, S1, F1, F2, and N1) requested future camera-based or sensor-based analysis that could provide \textit{performance-process information}.
Y1C1 requested a new motion-analysis function that could explain \textit{``what caused a high-ball foul''} or \textit{``which part of the defensive movement was insufficient''} when a player failed to stop the ball.
A high-ball foul occurs when the thrown ball does not make the required early contact with the court after release~\cite{ibsa2025goalballrules}.
N1P3 adapted a tactile match-viewing display encountered at the Paralympics into a proposal for individual throwing practice, allowing players to review the ball's course by touch after each throw.
Together, the proposals ranged from automatically announcing target-zone and foul results to explaining body movement through camera or sensor analysis and representing a throw's course through touch.

\subsubsection{Developing movement and spatial judgments through experience and external support}
\label{sec:findings-sensory-learning}
Across all seven teams, 16 players and 12 coaches collectively described three aspects of movement and spatial learning: combining external guidance with players' own sensory practice, variation associated with prior visual and sport experiences, and proposed technologies for individual throwing and defense practice.

% Evidence path: B07 and B09, SYN003, REL013, REL014, REL016, REL027, REL031.
% Supporting references: REF0026 (F1C1), REF0034 (N1P4), REF0138 (S2P3), REF0139 (Y2C3), REF0140 (Y1P1), REF0480 (F2P2), REF0636 (F2C2), REF0807 and REF0857 (Y1C1), REF0867 (F1C1), REF0874 (N1P1), and REF0958 and REF1020 (S1P3).
% Direct quotation: REF0034 (N1P4).
\textbf{External guidance and players' own sensory judgments.}
Fifteen participants across all seven teams described learning to handle the ball, orient themselves on the court, and perform movements through external guidance alongside practice using their own tactile, auditory, and proprioceptive judgments.
Five participants from four teams (Y1, Y2, S1, and S2) described beginner drills that familiarized players with the ball, raised court lines, and their own court areas before introducing basic defensive posture and throwing movements.
When visual demonstration was inaccessible, N1P4, who was blind from birth, described learning through direct touch: \textit{``Because I am completely blind, sometimes I touch the movement that the coach or the older players are doing, feel it with my hands, and then follow it. Or the coach directly changes my posture by touching and adjusting it.''}
F1C1 preferred verbal explanation and used hands-on guidance only when it was necessary and mutually agreed, while F2C2 explained that coaches could not physically guide a player during a fast throwing motion and therefore corrected the movement between repeated attempts.
Y1C1 described children who feared being hit by the ball and, when discussing a researcher-proposed ball-position feedback system, considered vibration shortly before the ball arrived a possible aid to adaptation.
Y1C1 also requested support beyond tactile floor markings to help players learn their court area through changes in their position and movement.
He emphasized combining this support with practice without it.
However, F2P2 emphasized that players needed an opportunity to form their own sensory judgment before receiving feedback. 
F1C1 and N1P1 similarly warned that continuous location feedback could create dependence or confusion because players needed to locate the ball and recover their position through sound and bodily practice during actual play.
Together, participants described touch and feedback as selective or temporary support, while practice without that support remained necessary for developing the sensory judgments used in play.

% Evidence path: B08, SYN003, REL023, REL024, REL025, REL026, REL088.
% Supporting references: REF0493 (Y1C1), REF0617 (F2P2).
\textbf{Prior visual and sport experiences.}
% Nineteen participants across six teams (Y1, Y2, S1, F1, F2, and N1) described how players' prior visual and sport experiences influenced their ability to form internal representations of movement and interpret nonvisual cues about distance and location on the court.
Nineteen participants across six teams (Y1, Y2, S1, F1, F2, and N1) described how they formed internal representations of movement and interpreted nonvisual cues about distance and location on the court. 
These accounts varied with participants' visual history and prior sport experience.
Y1C1 described players with acquired vision loss who drew on memories of visually experienced spaces and earlier sports to relate the sound of an approaching ball to its distance and to the area of the court they needed to cover in defense.
F2P2, who was blind from birth, described difficulty translating imagined movements into bodily action: \textit{``Also, because I was born with a visual impairment, I really struggle with mental rehearsal. ... I am always Superman in my head. ... But when I try it with my body, it never comes out that way.''}
% Together, these examples identified prior experience as one source of variation in how individual players formed internal representations, interpreted sound, and carried out intended movements.

% Evidence path: B10, SYN003, REL035, REL036,  REL037, REL038, REL039, REL040, REL093, REL105. + SYN007, REL034.
% Retrieval difficulties: REF0618 (S1P3) and REF0619 (Y1P1).
% Requested support: REF0914 (S1P1), REF0915 and REF0916 (S1P3).
% Basic and ball-delivered defense practice: REF0105 (S1C1).
% Human-delivered defense practice: REF0088 (S1P2), REF0089 (N1C2), REF0090 (N1P4), REF0671 (F2P1), REF0672 (S1P3), REF0690 (F1C1), REF0691 (F2C2), REF0709 (N1C1), REF0710 (N1P1), REF0711 (N1C3), and REF0907 (Y2C3).
% Participant-originated spatial-audio concepts: REF0756 and REF0757 (F2P2), and REF0758 and REF0785 (F2C2).
% Researcher-presented virtual-defense concept and participant evaluations:  REF0760 (Y1C1), REF0762 and REF0764 (F2C1), REF0774 and REF0790 (F1C1), REF0779 and REF0787 (N1P1), REF0780 (S1P4), and REF0782 (F2C2).
% Direct quotation: REF0782 (F2C2).
\textbf{Technical requirements for individual practice.}
Twenty-five participants across all seven teams described constraints on repeated individual throwing and defense practice and evaluated the potential uses and limitations of proposed technologies.
S1P3 and Y1P1 described difficulty locating the ball during independent practice once it stopped moving and no longer made a sound.
S1P1 requested barriers to keep balls from traveling too far, while S1P3 proposed a ball-return mechanism to reduce retrieval time and continue practicing throws.
Beyond basic posture drills, ball-based defense practice generally required a coach or teammate to repeatedly send balls toward the player.
Eleven participants from five teams (Y2, S1, F1, F2, and N1) described this dependence and identified limits in the number, consistency, or physical workload of repeated human throws.
F2C2 explained that the thrower's shoulder strain could limit the number of balls delivered even when defenders were able to continue practicing.
Against this background, eighteen participants across all seven teams requested a motorized ball-launching machine that could repeatedly deliver balls with varied speeds and bouncing trajectories without requiring another person to throw every ball.
F2C2 noted that machine noise could mask the bells inside the ball and separately proposed reproducing directional ball sounds through earphones, with the perceived direction adjusted to the player's court position and head orientation.
F2P2 proposed a separate seven-speaker setup at the goal that would allow center players to learn how sounds vary depending on where the ball lands.
% The research team asked four participants from three teams (S1, F1, and F2) to evaluate a virtual-defense concept that represented an incoming ball through spatial audio and vibration at the simulated point where the ball would contact the player's body.
The research team also presented a virtual-defense concept that represented an incoming ball through spatial audio and vibration at the simulated point where the ball would contact the player's body, and invited participants to discuss its potential uses and limitations.
F2C1 anticipated that centers could warm up independently with the system while coaches attended to wingers. 
Y1C1 expected the concept to sustain students' interest, improve parents' perceptions of goalball, and help broaden the player pool through use in schools.
N1P1 proposed recreational goalball facilities at welfare centers, community sports centers, or schools, where people with visual impairments could encounter the sport through play before progressing to recreational or competitive participation.
However, F2C2 explained why accurately reproducing an approaching ball sound would be difficult: \textit{``There are too many kinds of throws, and the sound also changes with the space. The ball sounds different in a narrow space and in a large space.''}
Two participants from two teams (N1 and F2) described contrasting spatial considerations: N1P1 emphasized the need for enough space to perform full-body defensive movements, whereas F2C1 considered the concept potentially useful when a full court was unavailable.
F1C1 wanted contact with a physical ball so that players could judge where it traveled after striking their body, while S1P4 noted that the simulation would not reproduce the transition from blocking to ball recovery and attack.
% Across these accounts, participants identified limits in reproducing ball sounds, physical contact, and the transition from blocking to recovery and attack, and anticipated opportunities for independent practice and wider engagement with goalball.
Across these accounts, participants identified opportunities to support ball retrieval, sustain independent practice, and broaden participation in goalball, alongside limitations in simulating ball sounds, physical contact, and the transition from blocking to recovery and attack.

% The next section examines how individual throwing and defensive skills were combined with coordinated movement, positioning, and communication in team practice.

\subsection{Team-Based Practice: Tactical Coordination and Communication}
\label{sec:findings-team-practice}
Building on the individual skills described in Section~\ref{sec:findings-individual-training}, team-based practice combined throwing and defensive actions with coordinated movement, positioning, and communication among three on-court players.
The following subsections describe coordinated attack, defense, and transitions in team practice, player communication during play, and tactical adjustment through coaching feedback and shared tactical information.

\subsubsection{Practicing coordinated attack, defense, and transitions}
\label{sec:findings-transitions}
Across six teams (Y1, Y2, S1, F1, F2, and N1), 11 players and 11 coaches collectively described three aspects of coordinated practice: scenario-based games and changing team combinations, moving attacks and the related positioning of teammates, and team defense connecting the initial block with ball recovery, passing, and the next attack.

% Evidence path: B21, REL042, REL043, REL044, and selected evidence from REL050 and REL061.
% Supporting references: REF0154 (F1P1), REF0156 (F2C1), REF0959 (F2C2), REF0157 (N1P4), REF0163 (F1P2), REF0516 (F1P1), and REF0219 (S1P3).
% Direct quotation: REF0154 (F1P1).
% REF0159 was excluded from REL043 following researcher review.
% REF0648 (F1P2) remains excluded from the Findings narrative.
\textbf{Scenario practice and team combinations.}
Twelve participants from four teams (S1, F1, F2, and N1) described practice games that used preset scores, restricted the types of throws, varied the number of players, assigned target areas, or changed team combinations to focus on particular situations.
F1P1 explained how these constraints changed three-on-three practice: \textit{``We might set a score, allow only a particular type of throw, or have one side defend with two players while the other three attack a particular player or course.''}
F2C1 and F2C2 described warm-up games without tactical direction from coaches, which allowed players to warm their bodies and try actions of their own.
N1P4 described changing combinations of players and coaching staff each week, and F1P2 described keeping the same player trio together for one week before changing teammates.
F1P1 noted that newly formed combinations required additional technical and tactical coordination, while S1P3 explained that unfamiliar teammates needed explicit calls for movements that familiar teammates often anticipated without speaking.
These formats used situational constraints and changes in team composition to focus practice on particular team responses rather than repeating the same full-match format.

% Evidence path: B22 and REL050.
% Supporting references: REF0189 and REF0263 (S1P3), REF0195 (N1C2), REF0214 (F1P1), REF0220 (N1P2), REF0221 (N1C3), and REF0230 (F2C1).
% Direct quotation: REF0230 (F2C1).
\textbf{Moving attacks and coordinated repositioning.}
Ten participants from six teams (Y1, Y2, S1, F1, F2, and N1) described moving attacks in which an attacker left the usual court position before throwing, changing the throwing location and making the attack harder for the opposing defense to predict.
% F1P1 described prearranged plays in which teammates moved together as an attacker crossed from one side of the court to throw from the other.
% N1C2 similarly described a wing player attacking from the opposite side or using space created when the center moved aside.
F1P1 and N1C2 described prearranged team movements that allowed an attacker to change throwing position while teammates created space.
F2C1 explained that his team repeatedly practiced these shared movement paths: \textit{``Because all three players enter the game with their eyes covered, if this movement has not been practiced enough, collisions can happen, injuries can happen, and performance can also decline.''}
Moving attacks were therefore practiced as coordinated team movements in which changing the throwing location depended on teammates creating space and following practiced paths safely.

% Evidence path: B23, SYN004, REL045, REL048, REL049, REL054, and REL064.
% Supporting references: REF0509 (F1C1), REF0526 (F1P1), REF0393 (S1C2), and REF0601 (Y2C3).
% Direct quotation: REF0526 (F1P1).
\textbf{Team defense, recovery, and transitions.}
Seventeen participants from six teams (Y1, Y2, S1, F1, F2, and N1) collectively described team defense as linking three-player positioning with the initial block, ball recovery, passing, and the next attack.
F1C1 explained that even when both wing players defended well individually, the defensive formation could still leave openings if the wing players and center did not maintain balanced spacing.
F1P1 explained, \textit{``Because three players have to defend the nine-meter-wide goal together, positioning and spacing among the three, and shared plans for backup defense when a ball deflects or rolls away, are very important.''}
S1C2 described drills that progressed from one defender to two and then three, then linked a deflected ball to recovery by the center, passing, and the next attack.
Y2C3 likewise characterized goalball as a defense-oriented sport in which a successful block had to connect to the next attack.
% Together, these descriptions showed that team defense depended on coordinated positioning as the ball approached and coordinated recovery, passing, and attack after the initial block.

\subsubsection{Player communication during team play}
\label{sec:findings-player-communication}
Across all seven teams, 14 players and 8 coaches collectively described three aspects of player communication: the center's coordinating role and workload, spoken and nonverbal sound cues used during play, and prearranged communication under rule-based time limits.

% Evidence path: B24, REL046, REL047, and selected center-related evidence from REL063.
% Supporting references: REF0200 and REF0201 (S2P2), REF0231 (N1P1), REF0499 (S1C2), REF0501 (S1P1), REF0503 and REF0642 (N1P4), and REF0604 (F2P2).
% Direct quotation: REF0231 (N1P1).
\textbf{The center's coordinating role.}
Fourteen participants from five teams (Y1, S1, S2, F2, and N1) described center players as spatial reference points and on-court coordinators.
S2P2 described calling out his position at center so that moving wing players could judge their direction and return to their positions.
N1P1 described how much communication the role required: \textit{``I think I am responsible for about 80\% of the sound in our games. I have to read and convey the entire flow of the game, including our attackers' ball paths and defensive timing.''}
N1P4 described the difficulty of learning different teammates' tendencies while also sharing tactical information and preventing collisions during moving attacks.
From a wing player's perspective, F2P2 characterized the center as carrying the team's greatest defensive burden because the position combined wide coverage, ball recovery, and distribution.
% Together, these descriptions showed that centers coordinated the team's spatial and tactical organization and that some center players experienced the concentration of these tasks as demanding.

% Evidence path: B25, REL048, REL052, and REL054.
% Supporting references: REF0183 (Y1P1), REF0193 (Y2C1), REF0210 and REF0235 (S1C2), REF0160 (N1C2), and REF0172 and REF0216 (F1P1 and F1P2).
\textbf{Spoken calls and nonverbal sound cues.}
Thirteen participants from six teams (Y1, Y2, S1, F1, F2, and N1) described how players used spoken calls and brief sounds to coordinate play, while also identifying limits to nonverbal communication.
S1C2 explained that defenders called out whether they had blocked or missed a ball, where it was rolling, and whether they could locate it.
N1C2 added that calling out when a blocked ball had deflected upward helped a teammate decide to provide backup defense.
For passes, S1C2 described the center calling the intended receiver's name, waiting for a reply that indicated the receiver's location, and then bouncing or rolling the ball so that its bells conveyed the path of the pass.
% Y1P1 described tapping the floor twice to signal his own location and clapping three times to ask a teammate to clear space for a moving attack, while Y2C1 described players using spoken location calls or ball taps.
% F1P1 and F1P2 described using a brief word or a cough to clear space during moving attacks, but they identified limits to relying on nonverbal signals because players could not see gestures under eyeshades and other court activity produced competing floor vibrations.
Y1P1 described using nonverbal sounds to communicate location and request space during moving attacks, while Y2C1 described spoken and nonverbal location cues.
F1P1 and F1P2 described brief spoken calls and nonverbal sounds for coordinating movement and space during moving attacks, but identified limits to nonverbal communication because players could not see gestures under eyeshades and other court activity produced competing floor vibrations.
The nonverbal cues described above included coughing, floor-tapping, and ball-tapping, but to protect team strategies, we do not specify which teams used each cue.

% Across these descriptions, teams combined concise speech with task-specific sound cues, but fast play and competing floor vibrations limited their reliance on nonverbal signals.

% Evidence path: B26, REL051, and REL053.
% Supporting references: REF0226 (F1P1, F1P2, and F1P3), REF0267 (F1C1), REF0165 (S2P2), REF0166 and REF0167 (Y1P1), REF0180 and REF0961 (S1C2), and REF0513 (F2P2).
% Direct quotation: REF0226 (F1P3).
\textbf{Prearranged communication under rule-based time limits.}
Eleven participants from six teams (Y1, S1, S2, F1, F2, and N1) described teams preparing concise words, codes, and time cues for two rule-bound situations: the brief intervals before the referee required silence and the 10-second period for making the next throw after the team first contacted the ball on defense~\cite{ibsa2025goalballrules}.
F1P3 emphasized the need for \textit{``a short expression whose meaning was immediately clear,''} while F1P2 described agreeing on short tactical terms before a match and F1C1 described rehearsing such terms before using them in competition.
% Y1P1 described the center varying the speed of floor taps to indicate whether the team should attack quickly or use more of the available time, while S2P2 counted elapsed time during ball recovery and then called the remaining seconds to a wing player.
Y1P1 described the center using auditory cues to coordinate attack tempo, while S2P2 counted elapsed time during ball recovery and then called the remaining seconds to a wing player.
S1C2 explained that when the center was occupied retrieving the ball, another player estimated and called the elapsed time.
F2P2 added that short, team-specific calls had to be rehearsed before competition because eyeshades, different venue conditions, and match pressure made in-match correction difficult.
Together, these descriptions showed that teams rehearsed concise tactical terms and divided timekeeping responsibilities so that they could coordinate the next action within rule-based time limits.

\subsubsection{Coach feedback and tactical adjustment during practice}
\label{sec:findings-coach-feedback}
Across six teams (Y1, Y2, S1, F1, F2, and N1), 6 players and 12 coaches collectively described how coaching feedback was timed and adjusted during practice and discussed technical requirements for making tactical information available to players and coaches.

% Evidence path: B28, REL056, REL057, REL059, REL072; B29, REL058.
% Supporting references: REF0244 and REF0528 (S1P4), REF0261 (F2C1), REF0199 (F1C1), REF0232 (F2P2), REF0246 (F2C2), REF0265 and REF0280 (F2C1).
\textbf{Timing and adjustment of coaching feedback.}
Fourteen participants from six teams (Y1, Y2, S1, F1, F2, and N1) collectively described how coaches managed the timing of feedback around play and adjusted it when an intended action was not carried out.
Coaches provided tactical feedback during brief pauses in play or after a practice game when repeated interruptions would disrupt practice.
S1P4 explained that coaching staff used a timeout to report where the ball had gone, substitutions, and changes in the opposing players' positions, while F2C1 described using pauses to redirect attacks after reviewing throwing courses.
F1C1 instead described waiting until a practice game ended before providing some corrections.
Within F2, F2P2 described a team agreement to shift a player's starting position when the same throwing-course error recurred.
F2C2 described giving feedback on how well a player carried out an assigned attack by considering both the player's movement before the throw and the throwing direction, while F2C1 described providing additional feedback, changing the starting position or lineup, and considering the player's physical or psychological condition when feedback was not readily accepted.
% Coaches therefore adjusted both when feedback was delivered and how they responded when execution did not improve.

% Evidence path: B30 and selected evidence from REL069, REL070, and REL071.
% Supporting references: REF0846 (N1C3), REF1018 and REF1019 (F2P2), REF0843 (F2C1), REF0917 (Y1C1), and REF0820 (Y2C1).
% Direct quotation: REF0846 (N1C3).
% REF0918 and REF0920 were excluded from the Findings narrative following researcher review.
% REF1019 was expanded in the source workbook.
% Its overlap with REF1018 was not counted as independent evidence.
\textbf{Technical requirements for tactical information.}
% Five participants from four teams (Y1, Y2, F2, and N1) requested or evaluated technologies that could track throwing courses, represent tactical locations through touch, show whether a three-player defensive formation matched the team's plan, or support learning and monitoring of the 10-second limit.
Five participants from four teams (Y1, Y2, F2, and N1) requested or considered technologies to support tactical analysis, spatial understanding, and time management within the 10-second limit.
N1C3 requested automatic overhead tracking of throwing courses during international competitions, explaining that \textit{``if that information could be available in real time, it would be much more helpful for developing or modifying tactics.''}
F2P2 proposed a proportionally scaled tactile model of the goalball court with raised court lines and movable magnetic markers, through which players could examine tactical positions and throwing paths by touch.
F2C1 requested video with a short playback delay so that coaching staff and players could verify whether the three-player defensive shift matched the team's planned coverage.
Y1C1 requested a system that could help players learn when to prepare and release a throw before the 10-second limit expired, while Y2C1 considered automatic monitoring of the 10-second limit a way for coaching staff to direct more attention to players.
% Together, these proposals and evaluations focused on making team-level tactical information available to players and coaches during training and competition.

% The following section examines how observations and records were reviewed after all training activities and used to inform subsequent training and competition preparation.

\subsection{Post-Training Activities: Performance Review, Analysis, and Feedback}
\label{sec:findings-review}
After training sessions and competitions, teams revisited video and performance records to review their own and opponents' play and inform subsequent practice and competition preparation.
The following subsections describe how teams collected and used these materials, how players and coaches addressed barriers to accessing and interpreting them, and which technologies participants requested to make post-activity analysis more accessible and efficient.

% Evidence path: SYN006, SYN008, REL074, REL075, REL076, REL080, REL081, and REL083.
% Supporting references: REF0312 (Y1P1), REF0305 and REF0967 (Y2C1), REF0311 (S1C1), REF0303 and REF0658 (S2P1), REF0300 and REF0301 (F1C1), REF0309 and REF0314 (F1P1), REF0319 (F1P2 and F1P1), REF0302 (F2C1 and F2C2), REF0307, REF0316, REF0539, and REF0656 (N1P4), REF0848 (N1C1), REF0654 (N1C2), and REF0293 (N1C3).
\subsubsection{Collecting and using video and performance records}
\label{sec:findings-review-practices}
Five players and eight coaches across all seven teams collectively described recording training sessions and matches, accessing archived match videos, reviewing the recordings individually or together, and supplementing them with manually produced performance records for subsequent training and competition preparation.
F2C1 and F2C2 described sharing recordings of training and matches through a private team link so that team members could watch them at home, while S1C1 explained that archived match broadcasts on YouTube had reduced the need for her team to record every match itself.
F1P1, who had acquired low vision, watched recorded training and match videos from his team's database on a mobile device and reviewed them alone or with teammates and coaches.
S2P1, who had congenital low vision, described reviewing recordings of team play with teammates, while coaches used the same videos to provide broader feedback on the play.
N1P4, who was blind from birth, described reviewing recorded practice with teammates or coaching staff through their verbal descriptions of her defensive posture, throwing form, and ball course.
Coaches from Y2, F1, and N1 described turning recorded matches into paper notes, spreadsheets, selected clips, or statistics on goals, fouls, and throwing courses.
For competition preparation, Y2C1 and N1C3 described analyzing videos of opponents' recent matches and using observed throwing and defensive patterns to plan simulations and formations for later team practice.
% Across these practices, recorded performance informed later training tasks and competition preparation.

% Evidence path: SYN009, REL077, REL078, REL079, and REL080.
% Supporting references: REF0299, REF0312, REF0537, REF0651, REF0886, REF0889, REF0996, and REF1011 (Y1P1), REF0308 (Y1C1), REF0289 and REF0538 (S1P3), REF0311 (S1C1), REF0306 and REF0658 (S2P2), REF0303 (S2P1), REF0310 and REF0881 (F1P1), REF0285 and REF0290 (F1C1), REF0657, REF0997, and REF0607 (F2P2), REF0307, REF0656, and REF0884 (N1P4), and REF0885 (N1P3).
\subsubsection{Access barriers and current adaptations}
\label{sec:findings-review-access}
Eight players and three coaches from six teams (Y1, S1, S2, F1, F2, and N1) collectively described different ways of accessing recorded performance and the remaining difficulty of interpreting visual material without assistance.
F1C1 described players with low vision watching enlarged recordings at close range and another person verbally describing the sequence of play for players who were blind.
% S2P2, who was also blind from birth, described using the announced moment of a conceded goal to recall what his body had felt during the recorded play.
F2P2, who was blind from birth, stated the central access barrier directly: \textit{``If I only have the video and no accompanying explanation, I can understand almost none of it.''}
He further perceived that athletes classified as B1, IBSA's lowest-vision sport class, were becoming less common among leading national teams, linking this pattern to informational advantages of residual vision, such as reviewing one's own throws without relying on others' explanations.
% \footnote{A manual tabulation of B1 designations in the official Goalball Results Books archived by IBSA identified 40/119 athletes at Rio 2016 (33.6\%), 33/118 at Tokyo 2020 (28.0\%), and 19/96 at Paris 2024 (19.8\%), indicating a successive decline in the proportion of B1 athletes across these three Games~\cite{ibsaHistoricalResults}.}
S1P3, who had acquired blindness, followed ball sounds but asked teammates with low vision to describe visual details such as the thrower's arm path, distance traveled, and number of steps.
S1P3 described difficulty locating a particular segment in a long recording, while Y1P1 reported that audio description was inconsistently available across archived matches.
Y1P1, who was blind from birth, used general-purpose artificial intelligence to examine photos and videos independently, but he found that it sometimes interpreted goalball movements through the conventions of other sports.
These descriptions showed that access depended not only on whether a recording existed but also on whether players could locate relevant moments and obtain goalball-specific information in an accessible form.

% Evidence path: SYN006, SYN009, SYN010, REL078, REL079, REL081, REL082, REL083, and REL084.
% Supporting references: REF0657 and REF0997 (F2P2), REF0879, REF0889, and REF1011 (Y1P1), REF0880 and REF0888 (S2P2), REF0881 (F1P1), REF0882 and REF0884 (N1P4), REF0883 (N1P3), REF0298, REF0301, REF0540, REF0932, REF0933--REF0938 (F1C1), REF0653 and REF0806 (N1C2), REF0655, REF0849, REF0931, and REF1009 (F2C1), REF0848 (N1C1), and REF0652 (N1C3).
% REL082 is represented here only by REF0653, REF0655, REF0806, and REF0849 because its other references concern in-session or individual-practice feedback.
\subsubsection{Technical requirements for accessible post-activity analysis}
\label{sec:findings-review-analysis}
Six players and five coaches from five teams (Y1, S2, F1, F2, and N1) collectively described technologies for making recorded performance accessible to players and expanding the analyses that teams could conduct after training or competition.
N1P4, who was blind from birth, requested a system running on a screen-reader-equipped computer that would privately communicate where the ball traveled and where goals were conceded through personal earphones or braille output.
N1P3, who had congenital low vision, proposed artificial intelligence that could analyze recorded video and provide spoken feedback about ball locations and events, including brief numerical codes.
Y1P1, who was blind from birth, requested spoken artificial-intelligence descriptions for online goalball video, while F1P1, who had acquired low vision, requested accessible software with goalball-specific knowledge that could convert match video into a form players could use without continuous interpretation by sighted staff.
S2P2, who was blind from birth, further emphasized that automated descriptions should be brief and timed so that the original ball and game sounds remained audible.
These proposals emphasized independent access to selected, goalball-specific information rather than simply adding more narration.

Coaches also requested technology that could reduce manual processing and produce analyses that current video did not readily support.
F1C1 estimated that manually extracting performance data from one match video could take a few hours, which led staff to select only the information they considered most necessary.
He expected automation to make it feasible to analyze more matches and requested searchable records by country, competition, and player.
He wanted concise summaries to support coaching feedback and accessible queries through which players could independently investigate opponents' patterns.
N1C3 described manually selecting attacking, defensive, scoring, and conceded-goal clips and tracing ball courses because these events could not be extracted automatically, while N1C1 requested automatic post-processing of each player's throwing-course frequencies and distributions.
N1C2 identified limitations in examining rapidly unfolding play through a single oblique camera view.
He argued that camera placement should match the intended analysis, with overhead views for team tactics and player-focused views for detailed throwing and defensive movements.
F2C1 requested automatic extraction of three-dimensional coordinates from recorded video because the vertical dimension of ball travel was missing from two-dimensional analysis.
% These requests focused post-activity analysis on reducing manual work, preserving the vertical dimension of ball travel, and making recurring performance patterns available for later review.
These requests focused on reducing manual processing, preserving the vertical dimension of ball travel, and making broader match records accessible for coaching feedback and players' own tactical questions.

\subsection{Training Organization: Operations, Resource Management, and Support Environment}
\label{sec:findings-operations}
Training across the preceding domains also depended on how teams organized shared programs, managed interpersonal and material resources, and assessed whether proposed technologies fit the physical and sensory conditions of goalball.

% Evidence path: SYN001, SYN003, SYN005, SYN011, and SYN012, REL085, REL086, REL087, REL088, and REL092.
% Supporting references: REF0328 (N1C1), REF0336 and REF0593 (F2C1), REF0361 and REF0678 (Y2C1), REF0392 (S1C1), REF0120 and REF0677 (S1P4), REF0666 (S1P3), REF0668 (Y1P1), REF0157 (N1P4), REF0163 (F1P2), REF0352 and REF0363 (F1C1), REF0354 and REF0355 (F2C1), REF0356 (Y1C1), REF0359 (F2P1), and REF0693 (S1P1), REF0554 and REF0803 (Y1C1), REF0552 (N1P1).
% Direct quotation: REF0120 (S1P4).
\subsubsection{Organizing training around player goals, schedules, and access needs}
\label{sec:findings-training-organization}
Eight players and six coaches from six teams (Y1, Y2, S1, F1, F2, and N1) described training goals and how shared sessions were organized around competition and daily schedules, attendance, and vision-related access needs.
Y1C1 described younger players losing interest without continuing feedback and sought game-like practice and device-delivered feedback to encourage participation. 
% Additional evidence: REF0577 (S1P3), linked through REL094.
S1P3 emphasized learning from higher-level teams and playing more friendly matches to develop skills.
% N1P1, who described herself as a senior national-team player, emphasized maintaining her existing performance rather than further improvement.
N1P1 emphasized maintaining her existing performance rather than further improvement.
Teams changed the balance of physical conditioning, basic skills, tactical practice, and simulated games according to competition schedules.
N1C1 described an annual progression from foundational work toward competition-specific training, while F2C1 and Y2C1 described increasing goalball-specific or tactical practice as competitions approached.
Daily schedules and access to the court narrowed what teams could do together.
Y2C1 explained that school schedules shortened the youth team's available training period.
At S1, a second-division team, work commitments among some players reduced attendance.
S1P4 added that physical conditioning was included, but \textit{``because many people share the training space for a limited period, it is difficult to devote much time to it.''}
F2C1 described a six-player roster that supported internal games, whereas S1P3 and Y1P1 described insufficient attendance that prevented team practice or resulted in reduced-player formats.
S1P1 further explained that a shortage of women players led women to practice with men.
% The activities possible in a shared session therefore reflected both the competition plan and whether enough players could use the court at the same time.
Within shared sessions, teams generally maintained a common program but adjusted how players accessed it according to vision condition.
F1C1 described providing blind players with additional orientation and repeated explanation before unfamiliar drills, while Y1C1 described using more detailed verbal explanation and physical guidance with blind than low-vision players.
F2C1 described adjusting players' positions and activity areas according to vision condition, and F2P1 described a lower treadmill speed as one individual adjustment for a blind player within shared conditioning.
Overall, training goals ranged from sustaining interest to maintaining performance, with sessions adapted to time, attendance, and vision-related access needs.

% Evidence path: SYN003, SYN007, SYN010, SYN011, SYN012, SYN013, and SYN014; REL093, REL095, REL096, REL098, REL099, and REL102.
% Supporting references for interpersonal challenges: REF0428, REF0431, REF0741, REF0939, and REF0940 (F1C1) and REF0559 (N1P1).
% Additional evidence: REF0689 (F2P2).
% Supporting references for injury management: REF0435 (N1C2), REF0436 (Y2C1), REF0440 (S1P3), REF0729 (N1C1), and REF0907 (Y2C3).
% Supporting references for facilities: REF0591 (S1C2), REF0702 (S1P2), REF0704 (Y2C1), REF0605 (Y1P1), REF0606, and REF0705 (F2C1).
% Supporting references for sport-science expertise: REF0398 and REF0579 (F1C1), REF0752 (Y2C2), REF0753 (S1P4), and REF0754 (Y2C1).
\subsubsection{Managing human, material, and external resources}
\label{sec:findings-resources}
Six players and eight coaches from six teams (Y1, Y2, S1, F1, F2, and N1) collectively described interpersonal challenges, coaching capacity, injury management, limitations of training facilities, and access to public support and goalball-specific expertise.
F1C1 explained that players could not see facial expressions that might clarify apology or effort, increasing the possibility of misunderstanding, and described using open discussion and individual conversations to address these situations.
F1C1 also described the burden of hearing different perspectives in interpersonal disputes.
He proposed a private messaging channel through which players who hesitated to speak up could contact their coach or other coaching staff and choose whether to continue by text or in person.
N1P1 similarly described how frustration could quickly disrupt verbal coordination when players performed under eyeshades.
F2P2 explained that coaching staff provided feedback on actions they had observed but sometimes could not answer his questions about a throw because they had been officiating or watching other players during practice games.
N1C2 and S1P3 described using finger taping and body padding, but N1C1 and Y2C3 still reported injuries among coaches and players who repeatedly threw balls during practice.
Y2C1 and S1P3 also described additional injury risk in mixed-gender practice when women and smaller players faced stronger throws from men.
The availability of training space mattered in both quantity and acoustic quality because players relied on the bells inside the ball.
S1C2 and S1P2 reported that uncontrolled noise in a shared open gym masked the ball sound, while Y2C1 and F2C1 described few dedicated courts and limited access to those available.
Y1P1 described limited public awareness of goalball, while F2C1 associated greater public interest with opportunities for advertising and sponsorship.
Y2C2 and S1P4 perceived a lack of goalball-specific scientific knowledge about movement and throwing techniques, while Y2C1 wanted evidence linking physical condition with performance.
% F1C1 described his team's response, which involved applying a new eight-week conditioning program developed with a national sport-science institute.
F1C1 described his team's response, which involved applying a new long-term conditioning program developed with a national sport-science institute.
Because the program had only recently begun, F1C1 considered it too early to assess its effects.
% Together, these cases showed that sustaining training required interpersonal and safety management as well as spaces and expertise suited to the sensory and physical demands of goalball.
These accounts highlighted the work of supporting players, managing injuries and interpersonal concerns, and securing facilities and expertise, alongside a need for greater public awareness.

% Evidence path: SYN007; REL036, REL103, and REL104.
% Supporting references for wearable safety: REF0829 (F1C1), REF0872 (N1P1), and REF0918 (S1C2).
% Supporting references for detection and tracking accuracy: REF0832 (S1C2) and REF0869 (N1C1).
% Supporting references for ball-launching machines: REF0908 (S1C1), REF0909, REF0913, and REF1014 (F2P2), and REF0910 (Y1P1).
\subsubsection{Design requirements for proposed training technologies}
\label{sec:findings-technology-constraints}
Three players and four coaches from five teams (Y1, S1, F1, F2, and N1) collectively identified physical safety, acoustic compatibility, precision, and goalball-specific ball control as requirements for proposed training technologies.
Because goalball players repeatedly dive and absorb ball impacts, participants questioned whether body-worn hardware could be used safely.
N1P1 warned that earphones or sensors could detach, break, or injure a player during defense.
F1C1 therefore called for earphones that remained in place and did not break on impact, while S1C2 raised similar concerns about wrist-worn vibrotactile feedback devices.
Automated systems also had to distinguish goalball-specific events precisely.
S1C2 explained that a foul-detection system would need to distinguish a ball crossing a line in the air from one contacting the floor.
% When the research team asked N1C1 to assess a researcher-proposed real-time location-tracking system using commercially available ultra-wideband (UWB) technology with an estimated positional error between 10 and 30 cm, the coach considered this range too large for goalball and suggested reducing the error to within 5 cm.
The research team asked N1C1 to assess a proposed real-time location-tracking system using commercially available ultra-wideband (UWB) technology, describing an estimated positional error of 10--30 cm~\cite{pozyxUwbRtls,sewioUwbAnchors}.
The coach considered this range too large for goalball and suggested reducing the error to within 5 cm.
The ball-launching machines raised several connected design considerations.
S1C1 emphasized speed control because contact with a fast-spinning ball could cause burns.
% Y1P1 explained that propelling the heavy ball required a powerful motor whose noise could interfere with listening to the bells inside the ball.
F2P2 added that the machine would need to control release angle, speed, course, and multiple bounce patterns, and anticipated that meeting these requirements would increase development cost.
Overall, participants treated physical safety, audible ball cues, event accuracy, and realistic ball delivery as prerequisites for development and adoption.

%% file: 5_Discussion.tex
\section{Discussion}
We discuss how goalball training practices, participants' experiences, and training challenges inform the design and evaluation of future support technologies.

% Evidence: REF0034 and REF0882 (N1P4), REF0285 and REF0290 (F1C1), REF0493 (Y1C1), REF0607, REF0617, REF0657, and REF0997 (F2P2), REF0844 (N1P3), and REF0880 (S2P2).
% The roster tabulation is researcher-provided context. Combining sensory feedback and evaluating how modalities complement one another are author-proposed research directions.
% \subsection{Eyeshades Equalized Play, but Vision Condition and History Still Shaped Learning}
\subsection{Accessible Learning Across Vision Conditions and Histories}
\label{sec:discussion-vision-learning}
%First, eyeshades equalized playing and training experiences in goalball, but vision condition and history of each player still shaped learning the required skills.
Goalball requires all players to wear eyeshades during play~\cite{ibsa2025goalballrules}, yet their current vision and visual histories still shaped how they developed movement skills through practice and understood their performance through review. 
Section~\ref{sec:findings-review-access} described enlarged video for players with low vision and verbal descriptions for blind players. 
Accounts in Section~\ref{sec:findings-sensory-learning} further described how players with acquired vision loss could draw on visual memories formed before sight loss, whereas players blind from birth learned without those visual references.
Although F2P2's account and the separately tabulated decline in B1 representation (Section~\ref{sec:findings-review-access}) do not establish a causal role for residual vision, they warrant closer attention to support for independent learning across visual status and history.\footnote{A manual tabulation of B1 designations in the official Goalball Results Books archived by IBSA identified 40/119 athletes at Rio 2016 (33.6\%), 33/118 at Tokyo 2020 (28.0\%), and 19/96 at Paris 2024 (19.8\%), indicating a successive decline in the proportion of B1 athletes across these three Games~\cite{ibsaHistoricalResults}.}
Participants requested tactile displays to examine their throws during practice and speech or braille output to review recorded play (Sections~\ref{sec:findings-process-feedback} and~\ref{sec:findings-review-analysis}). 
These needs point to further development of systems that combine auditory and tactile feedback across goalball training and review. 
Evaluation should examine how these modalities complement one another in helping players with different visual states and histories develop movement skills and understand performance.

% Evidence: REF0613 (N1P1), REF0614 (F2P1), REF0620, REF0629, and REF0689 (F2P2), REF0808 (Y1C1), REF0815 (S1P4), REF0844 (N1P3), and REF1008 (F1P1).
% Feedback presentation, comparison with intentions and bodily experience, and evaluation criteria are author-proposed directions.
\subsection{Process Feedback to Explain Performance and Guide Subsequent Practice}
\label{sec:discussion-feedback}
Under eyeshades, players could struggle to check whether their movement and the ball's path matched what they intended or felt.
Fast incoming throws also left little time to judge how to respond. 
Coaches sought to guide corrections, but monitoring multiple players sometimes left an attempt unobserved (Section~\ref{sec:findings-resources}). 
Even if the results of an attempt are known, it may still be uncertain why such results occurred. 
For example, repeatedly telling players that their throws went toward the middle did not explain why they missed the target (Section~\ref{sec:findings-process-feedback}).
Participants requested movement analysis to explain how to correct throwing fouls and unsuccessful defensive actions  (Section~\ref{sec:findings-process-feedback}). 
Alongside prior studies of offensive patterns and performance indicators~\cite{woo2018goalball,molik2015game}, these accounts highlight the information needed to turn an outcome into a possible adjustment.
Building on these requests, systems could provide brief outcome feedback after each attempt, with movement and trajectory details available on demand. 
Explanations should refer to the recorded attempt and distinguish observed movements from inferred causes, so players and coaches can compare them with the player's intentions and bodily experience.
% Evaluation should examine whether this support helps players understand how throws and defensive actions unfolded and decide what to change in subsequent practice.
However, participants expressed concern that continuous location guidance could interfere with developing the sensory judgments needed in play (Section~\ref{sec:findings-sensory-learning}).
Feedback should therefore match the training task and preserve opportunities for players to make these judgments independently.
With these considerations in mind, evaluation should examine whether this support helps players understand how throws and defensive actions unfolded and decide what to change in subsequent practice.

% Evidence: REF0214 and REF0526 (F1P1), REF0230 (F2C1), REF0509 (F1C1), REF0393 (S1C2), REF0601 (Y2C3), REF0200 and REF0201 (S2P2), REF0231 (N1P1), REF0503 and REF0642 (N1P4), and REF0603 and REF0604 (F2P2).
% The explanation for concentrated coordination work, training wings' spatial judgments, and evaluation of role sharing and centers' attacking opportunities are author-proposed interpretations and research directions.
\subsection{Spatial Training Methods to Help Wings Share Centers' Coordination Work}
\label{sec:discussion-coordination}
Under eyeshades, moving attacks required rehearsed routes to avoid collisions, while team defense depended on coordinated spacing, ball recovery, and the next attack (Section~\ref{sec:findings-transitions}). 
Much of this coordination fell to centers, some of whom described difficulty directing teammates alongside their own defensive responsibilities (Section~\ref{sec:findings-player-communication}). 
The wings' frequent changes of position offer one explanation for this concentration of work: the center's location calls provided a relatively stable reference for teammates finding their way during and after attacks. 
Sharing this work may therefore require wings to develop spatial judgments from their own changing positions. 
Technologies could support practice in locating teammates and judging available space during movement.
A related request concerned support for players to learn their court area as they moved and changed position (Section~\ref{sec:findings-sensory-learning}). 
However, evaluation should examine whether wings can use these skills to share coordination during attacks, defensive returns, and ball recovery, and whether this gives centers more opportunities to attack.

% Evidence: REF0554, REF0760, REF0803, REF0807, and REF0857 (Y1C1), REF0552, REF0559, and REF0787 (N1P1), REF0431, REF0741, and REF0932--REF0940 (F1C1), REF0691 (F2C2), REF0729 (N1C1), REF0907 (Y2C3), REF0606, REF0762, and REF0764 (F2C1), and REF0605 (Y1P1).
% Author interpretations and extensions concern continued participation, the significance of private communication for team coordination, broader practice and analysis opportunities, links from recreational experiences to actual teams, and evaluation of sustained involvement.
% Recreational facilities and progression toward organized participation were proposed by N1P1. Include REF0787 in Findings 4.1.3 to support this account.
\subsection{Support for Pursuing Team Goals Under Resource Constraints}
\label{sec:discussion-goals-resources}
% A youth coach considered temporary vibration cues useful for children afraid of being hit by approaching balls and requested playful feedback to sustain their interest (Sections~\ref{sec:findings-sensory-learning} and~\ref{sec:findings-training-organization}). 
% These proposals concerned willingness to begin and continue practice as well as skill development. 
% By contrast, a senior national-team player prioritized maintaining her existing performance (Section~\ref{sec:findings-training-organization}).
Across youth, second-division, and national-team accounts, training goals included sustaining participation, developing skills, and maintaining performance (Section~\ref{sec:findings-training-organization}).
% Rephrasing required
% Pursuing these goals also depended on the time and effort of coaching staff and teammates.
Pursuing these goals required support from coaches and teammates, access to goalball-specific training knowledge, and suitable training spaces, yet participants described constraints on each.

For solo throwing practice, goalball players prompted a request for automatic ball return because a stopped ball does not generate sound cues (Section~\ref{sec:findings-sensory-learning}).
Independent practice therefore required support for continuing repetitions as well as understanding each attempt.
For defense, repeated throwing strained coaches' and teammates' shoulders, motivating requests for ball launchers to sustain practice with less physical burden (Sections~\ref{sec:findings-sensory-learning} and~\ref{sec:findings-resources}).
Manual video analysis also consumed staff time and limited the information extracted, motivating automated, searchable records that could enable broader analysis (Section~\ref{sec:findings-review-analysis}).
Human support also involved addressing interpersonal tensions that could disrupt spoken cooperation, giving added importance to private channels through which players could raise concerns with coaching staff (Section~\ref{sec:findings-resources}).

Youth goalball coaches in prior research reported limited access to coaching manuals and requested education tailored to coaching young athletes with visual impairments~\cite{esatbeyouglu2024coaching}.
Our study makes these calls more concrete by identifying requests for coaching resources on knowledge linking players' physical condition to goalball performance, especially about producing varied throws and sport-science evidence (Section~\ref{sec:findings-resources}).
Such resources could help players and coaches make informed decisions about selecting and adjusting training methods.

Participants also saw virtual defense as a way to practice in limited space and warm up independently while coaches attended to other players (Section~\ref{sec:findings-sensory-learning}). 
A youth coach anticipated greater student interest and a broader player pool, while a national-team player proposed recreational goalball facilities as an introduction that could lead to organized participation. 
These expectations are relevant to concerns about goalball's limited public visibility (Section~\ref{sec:findings-resources}). 
Building on them, recreational booths, analogous to screen golf, could show visitors where and how to try goalball with a local team. 
Evaluation should examine whether these introductions lead to sustained participation and help recruit players or support personnel.

\textbf{Summary--} 
Prior studies have examined goalball experience within particular participant groups or settings~\cite{macdonald2020you,eddy2003mental,de2024coaching}, whereas our discussion examines how training experiences intersect with visual conditions and histories (Section~\ref{sec:discussion-vision-learning}), positional responsibilities (Section~\ref{sec:discussion-coordination}), and team contexts (Section~\ref{sec:discussion-goals-resources}) to shape support needs.
The resulting insights complement research on task-specific technologies~\cite{miura2018goalbaural,weber2015performance} by identifying design and evaluation considerations for balancing feedback with opportunities for independent sensory judgment (Section~\ref{sec:discussion-feedback}), sustaining repeated practice, and improving access to goalball-specific training knowledge (Section~\ref{sec:discussion-goals-resources}).

%% file: 6_Limitation.tex
\section{Limitations}
\label{sec:limitations}
Interviews and focus groups with 30 players and coaching staff across seven South Korean teams provided varied perspectives on goalball training, although the sample's geographical scope and uneven representation across team levels, roles, gender, and vision histories limit broader comparisons. 
The study also lacked longitudinal follow-up to track changes in teams' training practices and support needs over time.
% Despite the limitations, the accounts of practices, perceptions, and challenges with participants' technical requirements possibly informs the design directions for future systems. 
% Also, our observations and coded findings might guide the evaluation of proposed technologies in actual training settings.
Our findings offer a basis for subsequent development and evaluation, although the proposed systems' feasibility, effectiveness, and sustained use remain to be examined in actual training settings.

%% file: 7_Conclusion.tex
\section{Conclusion}
This study examined how technological support needs emerge from the everyday training practices and experiences of goalball players and coaches.
Across individual training, team-based practice, performance review, and training operations, our analysis showed how learning and support needs varied with vision conditions and histories, team roles, training goals, and available resources. 
These connections inform design directions for technologies that help players learn independently, understand outcomes and decide what to change next, and share coordination work with teammates. 
This study provides an empirical foundation for developing and evaluating accessible technologies grounded in everyday goalball training.

%% file: 99_BIBLIOGRAPHIES.bib
@String{Computer = "{IEEE} Computer" }

@String{Springer = "Springer-Verlag" }

@manual{bestebroer2025goalball,
  author  = {{John Bestebroer}},
  title   = {Goalball: Training Materials},
  year    = {2025},
  url     = {https://sportswithoutlimits.hk/wp-content/uploads/2025/07/goalball_-_training_materials.pdf},
  urldate = {2026-07-15}
}

@manual{ibsa2025goalballrules,
  author       = {{International Blind Sports Federation (IBSA)}},
  title        = {{IBSA Goalball Rules and Regulations 2025 (v1.0)}},
  year         = {2025},
  organization = {International Blind Sports Federation},
  url          = {https://www.goalball.de/wp-content/uploads/Final-2025-IBSA-Goalball-Rules-and-Regulations-1.pdf},
  urldate      = {2026-07-15}
}

@inproceedings{de2023understanding,
  title={Understanding challenges and opportunities in body movement education of people who are blind or have low vision},
  author={De Silva, Madhuka and Goodwin, Sarah and Holloway, Leona and Butler, Matthew},
  booktitle={Proceedings of the 25th International ACM SIGACCESS Conference on Computers and Accessibility},
  pages={1--19},
  year={2023}
}

@inproceedings{tanaka2026touching,
  title={Touching movement: 3d tactile poses for supporting blind people in learning body movements},
  author={Tanaka, Kengo and Wang, Xiyue and Takagi, Hironobu and Ochiai, Yoichi and Asakawa, Chieko},
  booktitle={Proceedings of the Twentieth International Conference on Tangible, Embedded, and Embodied Interaction},
  pages={1--19},
  year={2026}
}

@inproceedings{miura2018goalbaural,
  title={Goalbaural: A training application for goalball-related aural sense},
  author={Miura, Takahiro and Soga, Shimpei and Matsuo, Masaki and Sakajiri, Masatsugu and Onishi, Junji and Ono, Tsukasa},
  booktitle={Proceedings of the 9th Augmented Human International Conference},
  pages={1--5},
  year={2018}
}

@inproceedings{weber2015performance,
  title={Performance Analysis in Goalball: Semiautomatic specific software tools},
  author={Weber, Christoph and Link, Daniel},
  booktitle={Proceedings of the 10th International Symposium on Computer Science in Sports (ISCSS)},
  pages={157--160},
  year={2015},
  organization={Springer}
}

@inproceedings{tsutsui2024low,
  title={Low vision boxing: Participatory design of adaptive kickboxing experiences with low vision person},
  author={Tsutsui, Ayaka and Yamamoto, Kenta and Zhao, Yinan and Suzuki, Ippei and Tanaka, Kengo and Ochiai, Yoichi},
  booktitle={Proceedings of the 26th International ACM SIGACCESS Conference on Computers and Accessibility},
  pages={1--18},
  year={2024}
}

@inproceedings{khurana2021beyond,
  title={Beyond adaptive sports: challenges \& opportunities to improve accessibility and analytics},
  author={Khurana, Rushil and Wang, Ashley and Carrington, Patrick},
  booktitle={Proceedings of the 23rd International ACM SIGACCESS Conference on Computers and Accessibility},
  pages={1--11},
  year={2021}
}

@article{gale2013using,
  title={Using the framework method for the analysis of qualitative data in multi-disciplinary health research},
  author={Gale, Nicola K and Heath, Gemma and Cameron, Elaine and Rashid, Sabina and Redwood, Sabi},
  journal={BMC medical research methodology},
  volume={13},
  number={1},
  pages={117},
  year={2013},
  publisher={Springer}
}

@article{miles2013qualitative,
  title={Qualitative Data Analysis: A Methods Sourcebook, 3/E.},
  author={Miles, Matthew B and Huberman, A Michael and Salda{\~n}a, Johnny},
  year={2013},
  publisher={Sage publications}
}

@article{strobel2025hci,
  title={HCI, Disability and Sport: A Literature Review},
  author={Strobel, Lukas and Gerling, Kathrin},
  journal={ACM Transactions on Computer-Human Interaction},
  volume={32},
  number={3},
  pages={1--41},
  year={2025},
  publisher={ACM New York, NY}
}

@article{kim2023forblind,
  title={Exploration  of  Constraints  and  Support  Measures  for  Participati on  in Lifetime  Sports  for  the  Visually  Impaired},
  author={Kim, Hwan and Kim, Ji-Tae},
  journal={Journal of Korea Society for Wellness},
  pages={15--21},
  year={2023}
}

@article{choi2024female,
  title={Exploring Constraints on Sports Participation and Persistence of Women with Visual Impairments: A Grounded Theory Perspective},
  author={Choi, Seoeun and Lee, Daseul and Park, Jinseok},
  journal={Korean Journal of Adapted Physical Activity},
  volume={32},
  number={4},
  pages={1--22},
  year={2024},
  publisher={Korean Journal of Adapted Physical Activity}
}

@article{woo2018goalball,
  title={Notational Analysis of the Attack Types of Goalball Games},
  author={Woo, Sangyeon and Shim, Yun-Sik and Ban, Yoon-Chang and Kim, Homook},
  journal={Korean Journal of Adapted Physical Activity},
  pages={83--92},
  year={2018}
}

@article{molik2015game,
  title={Game performance evaluation in male goalball players},
  author={Molik, Bartosz and Morgulec-Adamowicz, Natalia and Kosmol, Andrzej and Perkowski, Krzysztof and Bednarczuk, Grzegorz and Skowro{\'n}ski, Waldemar and Gomez, Miguel Angel and Koc, Krzysztof and Rutkowska, Izabela and Szyman, Robert J},
  journal={Journal of human kinetics},
  volume={48},
  pages={43},
  year={2015}
}

@article{eddy2003mental,
  title={Mental imagery in athletes with visual impairments},
  author={Eddy, Kate AT and Mellalieu, Stephen D},
  journal={Adapted physical activity quarterly},
  volume={20},
  number={4},
  pages={347--368},
  year={2003},
  publisher={Human Kinetics, Inc.}
}

@phdthesis{antritsou2016socialisation,
  title={The socialisation and motivation for participation of visually impaired and blind athletes in the sport of goalball},
  author={Antritsou, Maria},
  year={2016},
  school={Leeds Beckett University}
}

@article{choi2016sports,
  title={Can Playing Sport Has an Opportunity to Overcome a Barrier? - Empirical Study of Goalball Rookie Athlehes -},
  author={Choi, Young-Lae and Yu, Seong-Jin and Choi, Gyu-Hyeon},
  journal={Korean Journal of Adapted Physical Activity},
  pages={117--131},
  year={2016}
}

@article{de2024coaching,
  title={Coaching Behaviors in Goalball: Perceptions of Athletes With Visual Impairments},
  author={De Luna-Vazquez, Wellington and Shapiro, Deborah R and Moreno, Anthony},
  journal={International Sport Coaching Journal},
  volume={1},
  number={aop},
  pages={1--11},
  year={2024},
  publisher={Human Kinetics}
}

@article{macdonald2020you,
  title={“You think differently after playing this sport”: experiences of collegiate goalball players},
  author={MacDonald, Cathy and Bryan, Rebecca and Lieberman, Lauren J and Foley, John T},
  journal={Recreational Sports Journal},
  volume={44},
  number={2},
  pages={139--148},
  year={2020},
  publisher={SAGE Publications Sage CA: Los Angeles, CA}
}

@article{mycock2020blind,
  title={'The blind leading the blind’-A reflection on coaching blind football},
  author={Mycock, David and Molnar, Gyozo},
  journal={European Journal of Adapted Physical Activity},
  volume={14},
  number={3},
  year={2020},
  publisher={European Federation of Adapted Physical Activity (EUFAPA)}
}

@article{ovstehage2025exploring,
  title={Exploring key factors associated with training quality among elite para-athletes. A qualitative study},
  author={{\O}vstehage, Vemund and Sandbakk, Silvana Bucher and Kocbach, Jan and Sandbakk, {\O}yvind},
  journal={Frontiers in Sports and Active Living},
  volume={7},
  pages={1561641},
  year={2025},
  publisher={Frontiers Media SA}
}

@article{ovstehage2026coaches,
  title={Coaches’ perspectives on training quality in elite para-sport: a qualitative study of coach--athlete collaboration},
  author={{\O}vstehage, Vemund and Kocbach, Jan and Severin, Cecilia and Skovereng, Knut and Bucher Sandbakk, Silvana and Sandbakk, {\O}yvind},
  journal={BMC Sports Science, Medicine and Rehabilitation},
  year={2026},
  publisher={Springer}
}

@article{fernandez2021sports,
  title={Sports preparation in goalball: a systematic review},
  author={Fern{\'a}ndez Mu{\~n}oz, M{\'o}nica Paulina and Campos Campos, Kevin and Luarte Rocha, Cristian and Castelli, Luis Felipe},
  journal={Pensar en Movimiento: Revista de ciencias del ejercicio y la salud},
  volume={19},
  number={1},
  pages={136--161},
  year={2021},
  publisher={nc-sa/4.0/}
}

@article{esatbeyouglu2024coaching,
  title={Coaching Practices and Experiences of Youth Goalball Coaches},
  author={Esatbeyo{\u{g}}lu, Ferhat},
  journal={Uluslararas{\i} Bozok Spor Bilimleri Dergisi},
  volume={5},
  number={3},
  pages={171--181},
  year={2024},
  publisher={Yozgat Bozok {\"U}niversitesi}
}

@article{lee2022goalball,
  title={Oral life history of Tae-soon Han, the main contributor to the development of goalball and sports for the visually impaired},
  author={Lee, Inkyung and Im, Hyow-Kyoung},
  journal={Korean Journal of Adapted Physical Activity},
  volume={30},
  number={4},
  pages={1--18},
  year={2022}
}

@article{link2018finding,
  title={Finding the gap: An empirical study of the most effective shots in elite goalball},
  author={Link, Daniel and Weber, Christoph},
  journal={PloS one},
  volume={13},
  number={4},
  pages={e0196679},
  year={2018},
  publisher={Public Library of Science San Francisco, CA USA}
}

@article{morato2018faster,
  title={Faster balls increase the probability of scoring a goal in female and male elite goalball},
  author={Morato, M{\'a}rcio Pereira and Menezes, Rafael Pombo and Fonseca, Sofia and da Cunha Furtado, Ot{\'a}vio Luis Piva},
  journal={Revista Brasileira de Ci{\^e}ncias do Esporte},
  volume={40},
  number={4},
  pages={427--434},
  year={2018},
  publisher={Elsevier}
}

@article{cote1999exploratory,
  title={An exploratory examination of the coaching behavior scale for sport},
  author={C{\^o}t{\'e}, Jean and Yardley, John and Hay, John and Sedgwick, W and Baker, J},
  journal={Avante},
  volume={5},
  number={3},
  pages={82--92},
  year={1999}
}

@article{fujita2021ball,
  title={Ball trajectories and the probability of scoring a goal in elite male goalball throws},
  author={Fujita, Rafael Akira and Furtado, Ot{\'a}vio Luis Piva da Cunha and Morato, M{\'a}rcio Pereira},
  journal={European Journal of Adapted Physical Activity},
  volume={14},
  number={1},
  year={2021}
}

@misc{pozyxUwbRtls,
  key         = {Pozyx UWB RTLS tracks anything ultra accurate},
  year        = {2026-09-09},
  title       = {Pozyx Robust UWB real-time location system},
  url         = {https://www.pozyx.io/products/pozyx-uwb-rtls},
  author         = {Pozyx}
}

@misc{sewioUwbAnchors,
  key         = {UWB RTLS Tags for Indoor Location Tracking - Sewio RTLS},
  year        = {2026-09-09},
  title       = {UWB Anchors for Indoor Location Tracking},
  url         = {https://www.sewio.net/uwb-anchors/},
  note         = {Sewio Networks s.r.o.}
}

@misc{ibsaHistoricalResults,
  key         = {IBSA | International Blind Sports},
  year        = {2026-09-07},
  title       = {Goalball Historical Results},
  url         = {https://ibsasport.org/sports/goalball/about/historical-results/},
  note         = {International Blind Sports Federation}
}

@misc{ibsaGoalballWomensRankings2026,
  key         = {06-IBSA-Goalball-Womens-World-Rankings-June-2026},
  year        = {2026-09-04},
  title       = {IBSA Goalball World Rankings - Women},
  url         = {https://admin.ibsasport.org/api/media/2026/08/06-IBSA-Goalball-Womens-World-Rankings-June-2026.pdf},
  note         = {International Blind Sports Federation}
}

@misc{ibsaGoalballMensRankings2026,
  key         = {06-IBSA-Goalball-Mens-World-Rankings-June-2026},
  year        = {2026-09-04},
  title       = {IBSA Goalball World Rankings - Men},
  url         = {https://ibsasport.org/api/media/2026/08/06-IBSA-Goalball-Mens-World-Rankings-June-2026.pdf},
  note         = {International Blind Sports Federation}
}

@misc{plaudai,
  key         = {Plaud.ai - The World's No.1 AI Note-taking Brand},
  year        = {2026-08-21},
  title       = {Be present. Capture what matters.},
  url         = {https://www.plaud.ai/},
  note         = {Plaud.ai US}
}

@misc{nvivo,
  key         = {NVivo | The best AI qualitative data anaylsis (QDA) software},
  year        = {2026-08-21},
  title       = {Ask more from your data with the depth and power of NVivo},
  url         = {https://lumivero.com/products/nvivo/},
  note         = {Lumivero}
}

@misc{charmaz2006constructing,
  title={Constructing grounded theory: A practical guide through qualitative analysis},
  author={Charmaz, Kathy},
  year={2006},
  publisher={Sage}
}
